\documentclass[a4paper,fleqn]{cas-dc}
\usepackage{amsmath,amssymb,amsfonts,amsthm}
\usepackage{graphicx}
\usepackage[numbers]{natbib}
\usepackage{arydshln}
\usepackage{sectionbreak}
\newdefinition{remark}{Remark}
\newdefinition{definition}{Definition}

\usepackage{tikz}
\usepackage{pgfplots}
\newcolumntype{P}[1]{>{\centering\arraybackslash}p{#1}}

\def\tsc#1{\csdef{#1}{\textsc{\lowercase{#1}}\xspace}}
\tsc{WGM}
\tsc{QE}
\tsc{EP}
\tsc{PMS}
\tsc{BEC}
\tsc{DE}

\begin{document}
\let\WriteBookmarks\relax
\def\floatpagepagefraction{1}
\def\textpagefraction{.001}
\shorttitle{A Systematic Survey of Attack Detection and Prevention in Connected and Autonomous Vehicles}
\shortauthors{Limbasiya et~al.}

\title [mode = title]{A Systematic Survey of Attack Detection and Prevention in Connected and Autonomous Vehicles}

\author[1]{Trupil~Limbasiya}\ead{limbasiyatrupil@gmail.com}

\author[2]{Ko~Zheng~Teng}\ead{zhengteng_ko@mymail.sutd.edu.sg}

\author[3]{Sudipta~Chattopadhyay}\ead{sudipta_chattopadhyay@sutd.edu.sg}

\author[4]{Jianying~Zhou}\ead{jianying_zhou@sutd.edu.sg}

\address[1]{iTrust, Singapore University of Technology and Design, Singapore}

\address[2]{Information Systems Technology and Design, Singapore University of Technology and Design, Singapore}

\address[3]{Information Systems Technology and Design, Singapore University of Technology and Design, Singapore}

\address[4]{iTrust and Information Systems Technology and Design, Singapore University of Technology and Design, Singapore}

\begin{abstract}
The number of Connected and Autonomous Vehicles (CAVs) is increasing rapidly in various smart transportation services and applications, considering many benefits to society, people, and the environment. Several research surveys for CAVs were conducted by primarily focusing on various security threats and vulnerabilities in the domain of CAVs to classify different types of attacks, impacts of attacks, attack features, cyber-risk, defense methodologies against attacks, and safety standards. However, the importance of attack detection and prevention approaches for CAVs has not been discussed extensively in the state-of-the-art surveys, and there is a clear gap in the existing literature on such methodologies to detect new and conventional threats and protect the CAV systems from unexpected hazards on the road. Some surveys have a limited discussion on Attacks Detection and Prevention Systems (ADPS), but such surveys provide only partial coverage of different types of ADPS for CAVs. Furthermore, there is a scope for discussing security, privacy, and efficiency challenges in ADPS that can give an overview of important security and performance attributes.

This survey paper, therefore, presents the significance of CAVs in the market, potential challenges in CAVs, key requirements of essential security and privacy properties, various capabilities of adversaries, possible attacks in CAVs, and performance evaluation parameters for ADPS. An extensive analysis is discussed of different ADPS categories for CAVs and state-of-the-art research works based on each ADPS category that gives the latest findings in this research domain. This survey also discusses crucial and open security research problems that are required to be focused on the secure deployment of CAVs in the market.
\end{abstract}


\maketitle

\section{Introduction}
Connected and Autonomous Vehicle (CAV) is an automotive entity configured with revolutionary technologies such as sensors, robotics, and complex software. It automatically executes different automotive system operations (like computation and communication) for Vehicle-to-Everything (V2X) communications and In-Vehicle Network (IVN) data transmission through wireless technology, i.e., Dedicated Short-Range Communications (DSRC), Long-Term Evolution (LTE, i.e., 5G/6G), or Wireless Fidelity (Wi-Fi). The integration of such modern technologies with Intelligent Transportation Systems (ITS) is a powerful tool that can gather meaningful information for data analytics and provide real-time information and effective services to the end users, thereby benefiting society, vehicle passengers, other people, industrialists, governments in the development of a sustainable world \cite{Kim2021CS}, \cite{Sun2021ITS}. Therefore, CAVs are widely practiced in various applications, i.e., advanced road safety, business and human productivity, traffic flow and congestion management, data-driven mobility, sustainability, and transport accessibility. Hence, CAVs provide new business opportunities through next-generation automotive applications and services.

CAVs are mainly developed to offer effective productivity while commuting on the road. Thus, the control of various CAV components is primarily managed by the vehicle rather than the driver. To make it more straightforward for the implementation purpose, different automation levels for a vehicle are categorized while considering the level of vehicle and driver controls. A range of these levels starts at level 0 (no automation) and ends at level 5 (full automation). They are classified as (i) \textit{level 0}: no driving automation, (ii) \textit{level 1}: driver assistance, (iii) \textit{level 2}: partial driving automation, (iv) \textit{level 3}: conditional driving automation, (v) \textit{level 4}: high driving automation, and (vi) \textit{level 5}: full driving automation. In this classification, the first three levels are categorized as the driving environment based on human driver monitoring, whereas the automated system monitors the driving environment in the other three levels \cite{SAEAutomationLevels2021}.

\begin{figure*}
\centering
{\includegraphics[width=0.8\textwidth]{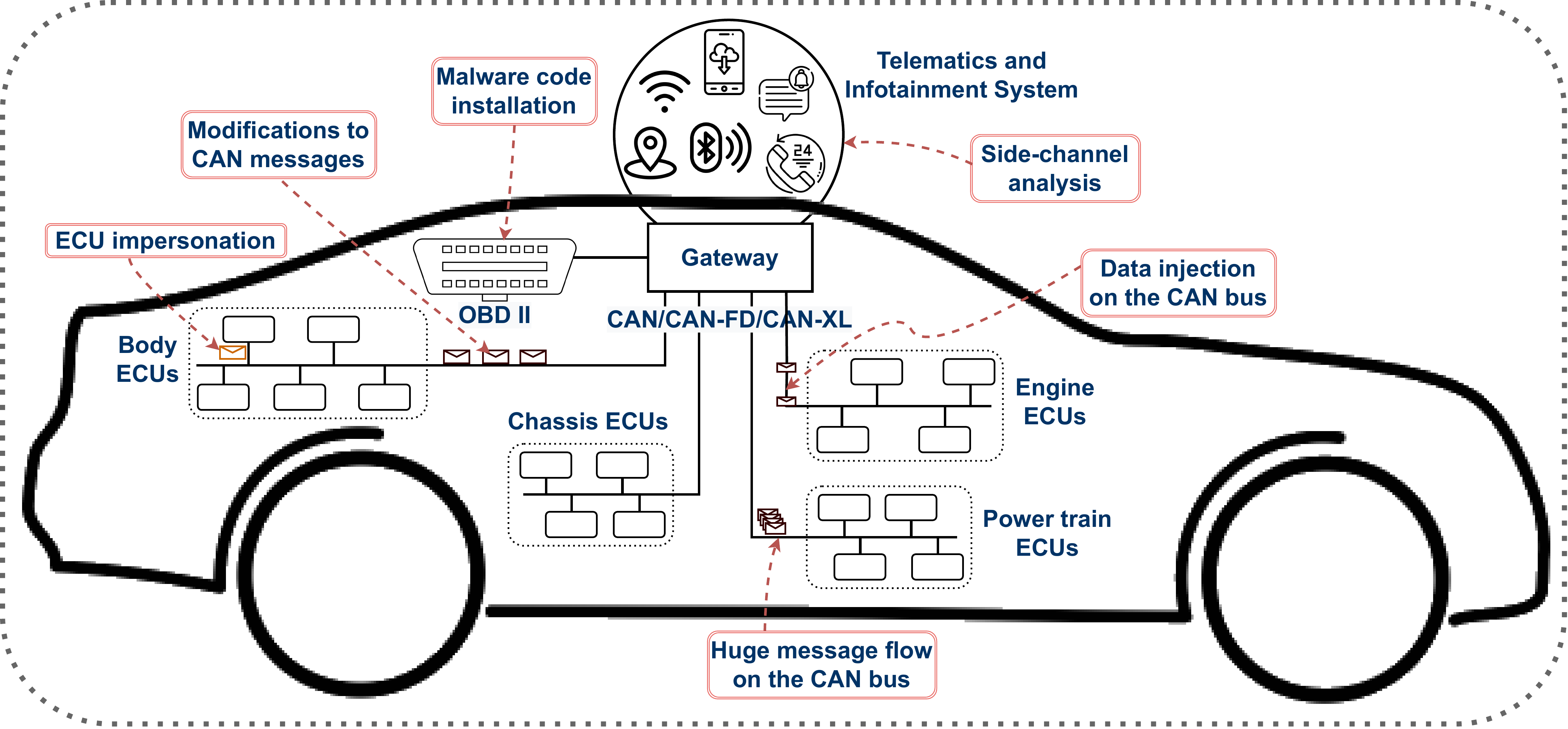}}
\caption{The Overview of Connected and Autonomous Vehicle with Different Components.}
\label{CAVOverview}
\end{figure*}

\subsection{Introduction of IVN and CAN}
IVN is the backbone of CAVs in today's modern vehicles for data computation and communication among different installed sensors and mechanical components within a vehicle \cite{Wang2018CST}. CAVs consist of several Electronic Control Units (ECUs) that are connected over the Controller Area Network (CAN) to transmit meaningful automotive instructions for further action(s). There are different communication protocols for IVNs, i.e., CAN, FlexRay, Media-Oriented System Transport (MOST), and Local Interconnect Network (LIN). Among these protocols, CAN is mainly practiced in an automotive network due to the effective data rate comparatively and bus topology to connect critical ECUs to a high-speed CAN bus and less-critical ECUs to a low-speed CAN bus for critical real-time data exchanges in the IVN. Such connections facilitate the quick broadcasting of crucial automotive messages with a higher priority. Moreover, high-speed CAN, i.e., ISO 11898-2 is particularly resistant to electrical interference and offers design flexibility while considering the cost of implementation. Moreover, the CAN protocol significantly manages arbitration and collision avoidance while messages are sent simultaneously, thereby solving the problem of message re-transmission~\cite{Zeng2016ICST}. Figure ~\ref{CAVOverview} displays the outline of a CAV \cite{Huang2017ISPEC}, \cite{Aliwa2021CSUR}, \cite{EASI2020NDSS} that connects with different types of ECUs, Telematics and Infotainment System (TIS) and On-Board Diagnostic (OBD) II through IVN to broadcast automated operational messages, whereas the outside world is connected via DSRC, LTE, or Wi-Fi for better services. Figure ~\ref{CAVOverview} also presents an outline of adversaries' target CAV components to perform illegal activities by launching impersonation, modification, injection, CAN bus-off, and side-channel attacks.

The CAN bus bit rate varies from 125 kbps to 1 Mbps with a payload up to 8 bytes, whereas the maximum bit rate for CAN Flexible Data (FD) is 8 Mbps, and the payload size is 64 bytes in CAN-FD. The third generation, CAN eXtra Long (CAN-XL) can provide a bit rate of up to 10 Mbps with a payload size of up to 2048 bytes, and it is implemented through Internet protocol-based services~\cite{CANXL2020}. CAN in CAVs is responsible for the overall behavior of different system functionalities, such as steering, engine management, braking system, navigation, lane/parking assistance, indicator panel, cruise control, and power window. Technological developments in recent years have allowed modern vehicles to access cloud services and to communicate with other vehicles using mobile cellular connections, thereby providing valuable services. However, they may also introduce new attack surfaces, leading to advanced security vulnerabilities that can be launched to disturb CAV components. Through the compromised ECU, the attacker can take control of the vehicle, which may result in severe consequences, e.g., the attacker can alter the speed of the vehicle or stop the vehicle altogether~\cite{Kim2021CS}, \cite{Jo2021ITS}.

\subsection{Market Scope of CAVs}
Automotive applications and services are used in various industries, such as transportation, retail, autonomous vehicles, financial, insurance, energy, health services, and media for multiple purposes that lead to a huge market scope of automotive businesses in the present and future world. According to a survey~\cite{FutureMobility2015Deloitte}, the automotive industry sector's total annual revenue in 2014 was around 2 trillion USD in the United States (US) only, which was 11.5\% of US Gross Domestic Product (GDP) in the year of 2014. Around USD 735 billion (of the total annual revenue) was explicitly generated from autonomous vehicles~\cite{FutureMobility2015Deloitte}. While looking at the roadmap of the New South Wales region of Australia~\cite{FutureTransport2021NSW}, CAVs will be adopted at a large scale in a service environment for different usages, aiming for new economic opportunities, great connectivity in customers' lives, and better accessibility of places through data analytics, new technologies, and strong collaborations. Moreover, autonomous vehicles provide functionality and services that are beneficial to decrease energy requirements and achieve sustainable mobility development. Such functionality and services include: (i) vehicle lightweight and rightsizing, (ii) powertrain electrification, (iii) platooning, and (iv) eco-driving~\cite{CAVEnergyConsumption2017EIA}.

Worldwide, e-commerce sales extensively grew to around USD 3.5 trillion from USD 572 billion in the period of 2010-2019~\cite{ECommerce2021Jilt}, and even more people have become E-commerce customers due to the Covid-19 pandemic for multiple individuals and societal benefits. The demand for last-mile delivery has therefore increased exponentially, resulting into higher delivery costs, longer delivery times, and fixed time slots due to limited human resources. Moreover, the environment will have negative impacts as delivery traffic continuously increases. Therefore, it is necessary to mitigate adverse effects for a sustainable development. To deal with these challenges, Autonomous Vehicles (AVs) can play a significant role in delivering various products to customers effectively and quickly to fulfill customer preferences, leading to a new delivery concept as Anything to Consumer (X2C). Considering the trend in shopping (for any products), the X2C delivery market will dominate regular parcel delivery in the near future that can be possible through Automated Guided Vehicles (AGVs) to deliver products in urban areas and Unmanned Aerial Vehicles (UAVs) for rural or hilly areas delivery, benefiting customers, businesses, and government~\cite{ParcelDelivery2016McKinsey}. Considering the significant intentions of government agencies, automotive industries, and researchers, many economic opportunities extensively open various ways to develop and commercialize new components and systems through future mobility technologies.

\subsection{Security, Privacy, and Efficiency Challenges}
The market of CAVs is exponentially increasing for different automotive services and applications due to various benefits of CAN bus system-based CAVs that integrate the outside world and the IVN for better real-time data analytics. For example, CAVs interact with different components (via an available central gateway in IVN), such as wireless sensors, other vehicles, network infrastructures, pedestrians, and other smart devices over LTE, DSRC, and Wi-Fi technologies for sustainable mobility. However, the nature of messages broadcast in CAN opens the opportunity for the attackers to penetrate the CAN for susceptible activities in the system. Moreover, CAN does not provide in-built authentication and encryption facilities to protect the system from potential security attacks. Thus, CAVs are vulnerable to many security threats in the exposure of IVNs to the remote attackers \cite{Aliwa2021CSUR}, \cite{EASI2020NDSS}, \cite{Woo2014ITS}, \cite{Petit2014ITS}, \cite{Humayed2017IoT}, \cite{Palaniswamy2020TIFS}. It is also demonstrated through experiments on a Jeep Cherokee that compromised electronic control units (ECUs) can be remotely accessed to broadcast forged or bogus messages on the CAN bus system~\cite{Checkoway2011USENIX}, \cite{Miller2015BlackHat}. The discussed experimental results~\cite{Cai2019BlackHat} revealed the possibility of security threats in different Bayerische Motoren Werke (BMW) car models, thereby the remote attacker can control the IVN of CAVs where various types of ECUs, OBD-II, and TIS are connected over the CAN bus. CAN is eventually susceptible to different security and privacy attacks due to the unavailability of encryption mechanisms and poor management of access control.

CAVs communicate with heterogeneous devices to deliver meaningful information timely and provide the best available content that enables decision-making systems to offer better efficiency and effective results. CAVs are also capable of gathering movement and location-based data of travelers through the installed sensors, and this captured data can be saved into the database (by using cloud-assisted systems) to analyze it based on available software, providing meaningful information to end users~\cite{Bloom2017SOUPS}, \cite{Joy2017ICCCN}, \cite{Fu2020IWC}. The automotive system in CAVs is also linked with different user accounts to provide relevant services effectually. However, the exposure of IVN also directly reveals the users' personal and confidential data (i.e., private conversations, user activities, vehicle locations, payment details, etc.). Since the design and implementation of privacy protection regulations are comprehensively pending for the collected data from CAVs, the collected data is shared among different stakeholders, such as the government, private industries (as the third-party service provider), and people \cite{Humayed2017IoT}, \cite{Lim2018Energies}. Consequently, privacy protection in CAVs is essential to avert the disclosure of identifiable information, vehicle tracking, and personal activities.

ECUs are resource-constrained components, and they are connected over the CAN bus to regularly broadcast messages for delivering different automotive instructions to the receiver entities so the automotive system can make better decisions in CAVs, where the human intervention is significantly less or null. However, the broadcast nature also increases the computation and communication overheads on the CAN bus and receiver end. Besides, security mechanisms need resources during the implementation stage to verify the authenticity and integrity of the sender and transferred messages~\cite{Woo2014ITS}, \cite{Palaniswamy2020TIFS}, \cite{Limbasiya2022ICDCN}. Therefore, it is vital to minimize the requirement of computational resources in attack detection and prevention solutions to quickly identify intrusions and provide protection against crucial threats in CAVs.

\subsection{Motivation towards a New Survey Article}
Several survey articles~\cite{Sun2021ITS}, \cite{Jo2021ITS}, \cite{Petit2014ITS}, \cite{Wu2019ITS}, \cite{Pham2021CS} on the cyber security of CAVs are presented by discussing various security threats and vulnerabilities in the domain of CAVs. Such articles have focused on the classification of attacks, attack features, impacts of attacks, cyber-risk, defense strategies against attacks, and safety standards. CAVs perform different IVN operations based on the automotive control system for effective services, but there is remarkably less or null human interference \cite{SAEAutomationLevels2021}. To offer effective productivity and comfortable journey to vehicle passengers, CAVs are connected with the outside world via DSRC, LTE, or Wi-Fi \cite{Sun2021ITS}. As a result, attackers can perform adversarial activities to disrupt IVN functionalities by launching different attacks remotely. This may lead to a major disaster on the road~\cite{Aliwa2021CSUR}, \cite{EASI2020NDSS}, \cite{Woo2014ITS}, \cite{Petit2014ITS}, \cite{Humayed2017IoT}. It is hence necessary to effectively identify vulnerable activities, detect defective components, protect the automotive system, and recover from unexpected situations to avert infrastructure and human life damages. 

An Intrusion Detection System (IDS) is a software-based procedure or tool to monitor the system/network to capture any adversarial incidents or activities that infringe the system's normal functionalities \cite{GarciaTeodoro2009CS}. Attack Detection and Prevention Systems (ADPSs) are software-based approaches, developed to detect anomalies in the system and protect it from malicious activities to continue its operations. Though it is vital to timely detect incidents and prevent the automotive system, the importance on attack detection and prevention approaches for CAVs has not been covered extensively in earlier surveys~\cite{Sun2021ITS}, \cite{Petit2014ITS}, \cite{Wu2019ITS}, \cite{Pham2021CS}. There are some surveys \cite{Kim2021CS}, \cite{Aliwa2021CSUR}, \cite{Jo2021ITS}, \cite{Wu2019ITS}, \cite{Lokman2019EJWCN} with a limited discussion on attacks detection and prevention, but all different types of ADPSs for CAVs are unexplored that can indeed improve the automotive system of CAVs. We have summarized the coverage (based on important subjects) of relevant recent survey articles in Table \ref{CovergaeSurveyTable} to understand the status of existing security survey articles in CAVs. We also notice that the significance of key requirements for CAVs (in terms of security, privacy, and efficiency) are not discussed substantially. This motivated us to extensively discuss all ADPS categories for CAVs that can help detect security problems in IVN and reduce the damage cost through CAVs on the road. Focusing on the aforementioned requirements, we write a detailed survey on different categories of ADPS and potential challenges in these ADPSs. 

\begin{table*}[!h]
\centering
\caption{Subject Coverage Comparison of Related Survey Articles on Connected/Autonomous Vehicles Security}
\label{CovergaeSurveyTable}
\scalebox{0.78}{		
\begin{tabular}{|l|c|c|c|c|c|c|c|c|c|c|}  \hline
\rowcolor{lightgray} \hspace{0.4in}\textbf{Subject} & \textbf{\cite{Kim2021CS}} & \textbf{\cite{Sun2021ITS}}  & \textbf{\cite{Aliwa2021CSUR}} & \textbf{\cite{Jo2021ITS}} & \textbf{\cite{Petit2014ITS}} & \textbf{\cite{Wu2019ITS}} & \textbf{\cite{Pham2021CS}} & \textbf{\cite{Lokman2019EJWCN}} & \textbf{This Paper} \\ \hline
IVN security and privacy properties & $\mathbb{AU}$; $\mathbb{PR}$ & $\mathbb{AU}$; $\mathbb{IN}$; $\mathbb{PR}$; & $\mathbb{AU}$; $\mathbb{CO}$; $\mathbb{IN}$; & $\mathbb{AU}$; & $\mathbb{AU}$; $\mathbb{PR}$ & $\mathbb{AU}$; & $\mathbb{AU}$; $\mathbb{IN}$; $\mathbb{PR}$; & --- & $\mathbb{AU}$; $\mathbb{AV}$; $\mathbb{CO}$; $\mathbb{IN}$; $\mathbb{PR}$;\\ \hline

Attack scenarios and feasible attacks & $\checkmark$ & $\checkmark$ &  $\checkmark$ & $\checkmark$ & $\checkmark$ & $\checkmark$ & $\checkmark$ & $\checkmark$ & $\checkmark$\\ \hline

ADPS categories overview & $\times$ & $\times$ & $\times$ & $\checkmark$ & $\times$ & $\checkmark$ & $\times$ & $\checkmark$ & $\checkmark$\\ \hline

Analysis of state-of-the-art ADPS methods & Partial & $\times$ & Partial & Partial & $\times$ & Partial & $\times$ & Partial & Extensive\\ \hline

ADPS performance evaluation parameters & $\times$ & $\times$ & $\times$ & $\times$ & $\times$ & $\checkmark$ & $\times$ & $\times$ & $\checkmark$\\ \hline

\end{tabular}}
$\mathbb{AU}:$ authentication; $\mathbb{AV}:$ availability; $\mathbb{CO}:$ confidentiality; $\mathbb{IN}:$ integrity; $\mathbb{PR}:$ privacy;
\end{table*}

\subsection{Organization of Paper}
The remaining part of the paper is organized as follows: Section 2 provides an overview of important security and privacy properties, attack scenarios, possible attacks in CAV, and performance evaluation parameters for ADPS. Section 3 described the considered approach to include the most relevant articles in preparing a survey paper. Section 4 discusses the overview of different ADPS categories and state-of-the-art research works based on each ADPS category that gives the latest findings in this research domain. In Section 5, we suggest important and open security research problems that are required to be focused on for novel contributions in CAVs. Section 6 gives our concluding remarks on the survey article and intended fuzzing approaches for intrusion detection.

\section{Key Requirements for CAVs: Security, Privacy, and Efficiency}
This section discusses essential security and privacy properties for IVN. We explain crucial security and privacy threats that can significantly impact the IVN. We also discuss various malevolent ways, used for adversarial activities in IVN. Therefore, it may lead to security and privacy challenges that need to be addressed to protect the IVN from illegal actions. Moreover, different performance parameters are explained that are useful to measure the performance efficiency of CAN-based ADPS and to understand the performance results of future ADPS universally.

\subsection{Security and Privacy Properties}
We explain relevant security and privacy properties that are more important in IVN as the exposures of private data and the system may lead to various issues in CAVs \cite{Frassinelli2020IEEESP}.

\subsubsection{Authentication}
When messages are exchanged over a common communication channel, the receiver entity should confirm the sender and data exactness of transferred messages to prevent misleading information and forgery. Furthermore, confirming both entities (sender and receiver) through mutual authentication and key agreement mechanisms in two-way message communications that confirm data exchanges between legitimate entities is necessary. Otherwise, it may lead to impersonation and data modification attacks, resulting into infrastructure damage and/or life threats to vehicle travelers. ECUs are connected over the CAN bus to send relevant messages to execute different operations in the automotive systems, and it thus becomes necessary to authenticate the sender in CAVs to avoid counterfeit information. If the sender is not authenticated, then there is a possibility that adversaries can perform malicious actions to interrupt IVN functionality, aiming to damage the automotive system in CAVs \cite{Kim2021CS}, \cite{Sun2021ITS}, \cite{Aliwa2021CSUR}, \cite{Jo2021ITS}, \cite{Petit2014ITS}, \cite{Wu2019ITS}, \cite{Pham2021CS}. To satisfy authenticity in the CAN, various security solutions that are mainly designed using MACs and digital signatures can be practiced.

\subsubsection{Availability}
It refers to the reliability of obtained information at the receiver side within a stipulated time to consider as the input in further actions. If imperative information is unavailable to the authorized entity at the required time, it may lead to unfortunate events that can put the entire automotive system in impairment situations \cite{Qayyum2020IEEECOMST}. CAVs are configured with the CAN to perform different automated operations based on the collected/given vehicular data through IVN functionality. Crucial automotive components (i.e., engine, power train, tire pressure monitoring, etc.) should receive instructions without any delay to execute related operations successively for providing an impeccable experience in autonomous vehicles. If exigent data is unavailable to crucial automotive components, it may lead to vehicle accidents on the road that might also have direct risks to human life. Moreover, it is also required to send appropriate messages if data delivery fails for some reason. Therefore, data availability is a crucial requirement in CAN-based CAVs.

\subsubsection{Confidentiality}
When messages are transferred over a public channel, it is essential to ensure that only legitimate receivers understand such information from the sent messages to satisfy the secrecy of data. Data confidentiality is lost if adversaries or other entities can extract meaningful information by intercepting exchanged messages. A compromised device can then disclose confidential information while sending data into the system. IVN messages are sent over the CAN bus and include automotive instructions used as input to perform further operations. Further, transmitted messages are available to all connected ECUs due to the broadcast nature of the CAN. Thereby, suppose any compromised ECU is connected over the CAN bus (and there is no guarantee that other ECUs know the connectivity of compromised ECUs over the CAN bus). In that case, an adversary can use confidential information of CAVs during malevolent activities \cite{Aliwa2021CSUR}. Thus, encryption of CAN messages has become essential during data transmission, but it is also vital to consider the resource-constraint problem in CAN while applying encryption techniques to satisfy data confidentiality.

\subsubsection{Integrity}
When messages are transferred from the sender to the receiver side, information should be available at the receiver entity the same as the sender sent it. If the receiver obtains modified information, then transmitted messages are tampered, resulting in the loss of data integrity. Thereby, the receiver should discard such altered messages without considering them. If the receiver accepts amended information, it may lead to different decisions, as the received information is used as input values for further operations. The functionalities in CAVs are operated based on automatic operations, and the intervention of humans is highly less or completely null in an automated system-monitored driving environment. Therefore, data integrity requirement becomes more significant in CAVs to verify the exactness of obtained messages from various senders. If adversaries can alter the CAN message, it creates data integrity problems in the system. It is also challenging to identify the sender in CAN to report malicious actions (performed by the specific entity) due to the unavailability of sender information in CAN messages \cite{Sun2021ITS}, \cite{Aliwa2021CSUR}, \cite{Pham2021CS}. One-way hash-based authentication techniques are effective to avoid data tampering in CAN messages. Consequently, integrity is also vital in knowing whether exchange messages are altered or not.

\subsubsection{User Privacy}
CAN is highly useful in automotive systems to execute various automated operations in CAVs, UAVs, AGVs, the health sector, and other related application areas, reducing/removing human monitoring for rich experience during the usage. In such applications, the system is connected to different devices for effective data analytics by exchanging meaningful communications. Furthermore, data is particularly crucial, so it has substantial inhibitive impacts on the system and its users if it does not have adequate data protection mechanisms. Thus, if data leakage is possible in CAVs, it can expose vehicle users' activities, previously visited places, vehicle movements, and related actions \cite{Sun2021ITS}, \cite{Petit2014ITS}, \cite{Pham2021CS}. In order to avert illegal data access, it is required to ensure that only legitimate entities should know vehicle users' activities. Therefore, it is necessary to satisfy user privacy in CAVs.

\subsection{Attacks Vector}
The attack vector is the way to enter the system to launch a diversity of attacks. We describe different possible ways for adversarial activities in CAVs based on the CAN bus architecture, OBD-II, wireless interface, physical ports, and ECUs components \cite{Aliwa2021CSUR}, \cite{EASI2020NDSS}, \cite{Jo2021ITS}, \cite{Humayed2017IoT}, \cite{Pham2021CS}.

\subsubsection{External and Internal Adversaries}
The external adversary is an outsider entity that is capable to receive public channel parameters (transferred over the TIS). However, the system parameters (given to the registered entities during the registration/initial phase) are not available to such adversaries as the external adversary is not registered with the system for legally executing various operations and communications in the future. Therefore, this type of adversary has limited capabilities to launch various attacks on the IVN components of CAVs. However, external adversaries can monitor exchanged messages (transferred over the public communication channel) to eavesdrop on such messages and then use such information to intercept future communications through adversarial actions.

An internal adversary (also known as an insider attacker) can be an authorized entity (who is registered with the central authority) to communicate with other registered entities. Thus, an internal adversary has own credentials, system parameters, and public channel parameters, enabling the adversary to perform malicious activities (launching diverse attacks) on the IVN components, i.e., ECU, CAN bus network, and infotainment system. Since internal adversaries aim to execute illegal actions on the system as a registered entity to avoid identification and detection. It is indeed challenging to identify inside attackers.

\subsubsection{Active and Passive Adversaries}
An active adversary aims to interrupt prevailing functionalities of the AV systems by generating/sending packets with deceitful intents to the system. Thereby, it has a direct impact on the AV systems which may persist important components of CAVs (i.e., engine/power train/chassis ECUs, CAN bus, gateway ECU) to perform abnormal actions, e.g., stop/delay messages, do changes in communications, overwhelm CAN packets, etc. Since CAVs enabled with the automation level 3/4/5 have very less or no human intervention, it becomes more crucial to manage CAVs during unexpected events (due to adversarial activities on the IVN). Therefore, malicious actions by active adversaries are more precarious comparatively that may lead to massive damage to the system infrastructure and/or humans.

A passive adversary mainly targets to eavesdrop on exchanged messages over the CAN bus and TIS to capture/learn meaningful information/communications from the collected data. An adversary then uses such eavesdropped data later for illegal purposes (i.e., forge message communications, modify messages, impersonate data transmissions, etc.) to disturb the IVN components of CAVs. This type of adversary is not easily identifiable as passive adversaries do not intercept the AV systems' functionalities directly.

\subsubsection{Local and Remote Adversaries}
When the manufacturers/service providers install tampered/vulnerable hardware equipment during the device installation or upgrading procedures, adversaries have an opportunity to connect with the automotive system devices (i.e., OBD-II, TIS, and ECUs) to understand on-going operations and other functionalities. Since such entities are overtly involved with the direct access of physical devices, they play a crucial role in the initialization/upgradation. An adversary can thus become more powerful to covertly perform harmful actions in the system, making it difficult to identify such compromises. The risk level in CAVs is therefore disastrous in the presence of compromised hardware components. Hence, it is important to protect CAVs from such local entities to minimize the system exposure.

Malicious code can be implemented in CAV equipments remotely to create backdoor vulnerabilities. This, in turn, enables the adversary to gain unauthorized remote control of the system/device \cite{Zhang2014IoTMalware}, \cite{Elkhail2021Access}. The remote attacker can then give commands through malware to execute damaging operations in CAVs and extend the scope of affecting areas through various malware vulnerabilities. Thus, adversaries can disrupt the functionalities of the CAN bus system architecture, software programs, and hardware devices/designs.

\subsection{Possible Attacks in CAV}
The conventional CAN and CAN-FD bus architectures are mostly used for real-time IVN communications due to the reduced cost, better efficiency, and simplified installation. However, they are exposed to various security attacks due to system vulnerability possibilities through the common CAN/CAN-FD communication bus, lack of authentication mechanisms, data encryption methods, and wide network connectivity over DSRC, Wi-Fi, LTE, and Bluetooth. As a result, the adversary can launch a variety of attacks over the CAN bus, ECUs, OBD-II, and keyless entry systems \cite{Aliwa2021CSUR}, \cite{EASI2020NDSS}, \cite{Jo2021ITS}, \cite{Humayed2017IoT}, \cite{Pham2021CS}. We have considered important attacks based on the significant impact on the IVN. Thereby, the explanation of pivotal attacks is limited to the CAN bus architecture, OBD-II, and ECUs, but we have not considered Light Detection And Ranging (LiDAR), Radio Detection and Ranging (Radar), and Global Positioning System (GPS)-based attacks in the following discussion. 

\subsubsection{Impersonation}
When the adversary gets the CAN bus network access, s/he can obtain all transferred messages due to the broadcast nature. Adversaries can learn the way of ECU behavior, i.e., CAN identity (ID), transmission rate, and payload range. The sender's information is not involved in CAN messages, making it easier to imitate ECU behavior by including the same information with an identical frequency. However, there is a possibility of a Denial of Service (DoS) attack due to the increment in the number of CAN messages, but an adversary can appropriately manage the timings for sending data over the CAN bus and disabling a particular ECU to launch an impersonation attack  \cite{Sun2021ITS}, \cite{Aliwa2021CSUR}.

\subsubsection{Modification/Fabrication}
This attack is performed to alter the CAN message payload with bogus information as CAN is enabled with the limited security features \cite{EASI2020NDSS}. The erroneous data is then sent to the receiver ECUs for misleading and to perform faulty operations in CAVs. Attackers can get a CAN ID of exchanged messages through the CAN bus connectivity, as well as the authentication and integrity properties are not implemented effectively in the CAN bus protocol. Therefore, adversaries can easily launch a modification attack to deliver fallacious information to disrupt vehicle functionalities by obstructing ECUs, TIS, and the CAN bus system. Since the adversary can broadcast incorrect data by injecting malicious messages, it is also difficult to correctly identify a modification attack due to a small amount of payload in CAN messages \cite{Sun2021ITS}, \cite{Aliwa2021CSUR}, \cite{Pham2021CS}.

\subsubsection{Sybil}
CAN does not offer robust security features, resulting into the exposure of IVN to the attackers \cite{EASI2020NDSS}. Authentication mechanisms are thus developed to protect illegal activities over the CAN bus architecture while exchanging CAN messages. However, the system requires computational resources to complete all the verification procedures before proceeding with the received data. All ECUs are connected over the CAN bus to continuously broadcast messages, but such ECUs are resource-constrained in nature \cite{Woo2014ITS}, \cite{Palaniswamy2020TIFS}, \cite{Limbasiya2022ICDCN}. Thus, authentication schemes should be cost-effective; otherwise, they may require more computational resources. Moreover, the receiver ECU takes more time to confirm the legality of received CAN messages, and it is overburdened with many remaining messages (which are required to be verified), including high-priority data, leading to a Sybil attack in IVN. Therefore, minimizing the impact of a Sybil attack is important by designing lightweight authentication protocols for CAN.

\subsubsection{Replay}
Messages are sent over the common CAN bus and are accessible to all the connected ECUs due to the broadcast nature of data transmission. The purpose of a replay attack is to stop transferred messages and re-transmit them (with or without modifications in data payload) later. Thus, the receiver ECUs cannot receive essential data timely, impacting vehicle services and functionalities provided through an automated vehicle system. Since CAN messages include crucial and real-time information which are used as an input in other operations to perform further executions, the unavailability of CAN messages to the receiver ECUs significantly impacts the overall IVN system. Besides, when adversaries broadcast CAN messages (already captured from the CAN bus) again, it increases the communication overhead on the CAN bus and the computation cost (to authenticate delayed messages) at the receiver ECUs. CAN components thus require more computational resources though ECUs have limited computing power to execute various operations \cite{Kim2021CS}, \cite{Sun2021ITS}. Hence, detecting a replay attack quickly is indispensable, reducing the additional requirement of computational resources and delivering CAN messages timely to make accurate decisions in CAVs.

\subsubsection{Injection}
An adversary aims to change the sequence of legal CAN frames, message frequency, the number of CAN frames for transmission on the bus, and message payload through an injection attack. Since CAN has limited security features to support authentication and encryption for transferred messages, adversaries can inject payload over the CAN bus (to fabricate messages) at an abnormal rate with unusual CAN traffic. This situation leads to the generation of simulated events that direct CAV parts to implement automotive operations based on the given instructions by an adversary. Thus, it directly interrupts the execution of crucial operations in the CAN \cite{Kim2021CS}, \cite{Aliwa2021CSUR}, \cite{Jo2021ITS}. Therefore, adequate authentication and integrity verification mechanisms are required to confirm exchanged CAN messages' legitimacy and exactness to protect from such attacks.

\subsubsection{CAN Bus-off}
Connected components to the CAN bus use the arbitration field to find the preference of CAN messages and then decide the occupancy of the CAN bus for delivering priority data first to the receiver ECU(s) before sending less important data to the destination. Since CAN messages are mainly sent without encryption (due to the fixed size of data fields), and a robust authentication mechanism is not available for CAN communications, adversaries can send many CAN messages with the highest attribution identity to dominate the CAN bus. Therefore, the communication link becomes unavailable to deliver crucial CAN messages for legitimate CAV components. Besides, the adversary can send the same CAN messages with a high frequency to overwhelm CAN resources to make the CAN bus unavailable for normal operations and functionalities of CAVs \cite{Kim2021CS}, \cite{Jo2021ITS}.

\subsubsection{Side-channel}
Adversaries can easily connect with the TIS applications and services and collect pertinent information (e.g., timing, energy consumption, cache, etc.) through connected devices (over wireless connectivity) with the IVN. In CAVs, in-vehicle infotainment systems can provide information, i.e., location, speed, access to related applications, and other data to make better decisions. Adversaries first analyze target systems' activities based on the collected pertinent information and then launch an attack on a specific system/user to disrupt normal operations and functionalities of the CAN. The target system is then exploited through analyzed data \cite{Li2019WCSideChannel}. Since the gathered data (based on various TIS applications and services) may include personal information, visited location, and other activities of vehicle users, it may lead to user privacy issues. Hence, protecting the IVN and avert from potential side-channel attacks is also important.

\subsubsection{Remote Sensor}
Sensors support various functionalities to measure environmental details, detect objects, and share information to effectively perform automotive operations in CAVs. Such delivered information is crucial in IVN as it is used as input to make different decisions. There are vital sensors, i.e., camera, ultrasonic radar, LIDAR, vision, sonar, and GPS \cite{Sikder2021CSTSensorSurvey}, \cite{Ma2020JAS}, which are used in crucial applications of CAVs. When inaccurate values are measured and not sent correct information to the receivers by the installed sensors, the automotive system may lead to spoofing and eavesdropping attacks \cite{Kim2021CS}, \cite{Sun2021ITS}, \cite{Aliwa2021CSUR}. Since CAN has limited security features, and demonstrations confirmed that adversaries could remotely access the CAN \cite{EASI2020NDSS}, security risks are significant in the automotive systems as CAVs have very less or no human intervention. Hence, t is essential to regularly check the status of sensors (in terms of faulty, compromised, or damaged) through intrusion detection mechanisms and protect data exchanges between such sensors and other IVN devices with encryption schemes.

\begin{figure*}
\centering
{\includegraphics[width=0.9\textwidth]{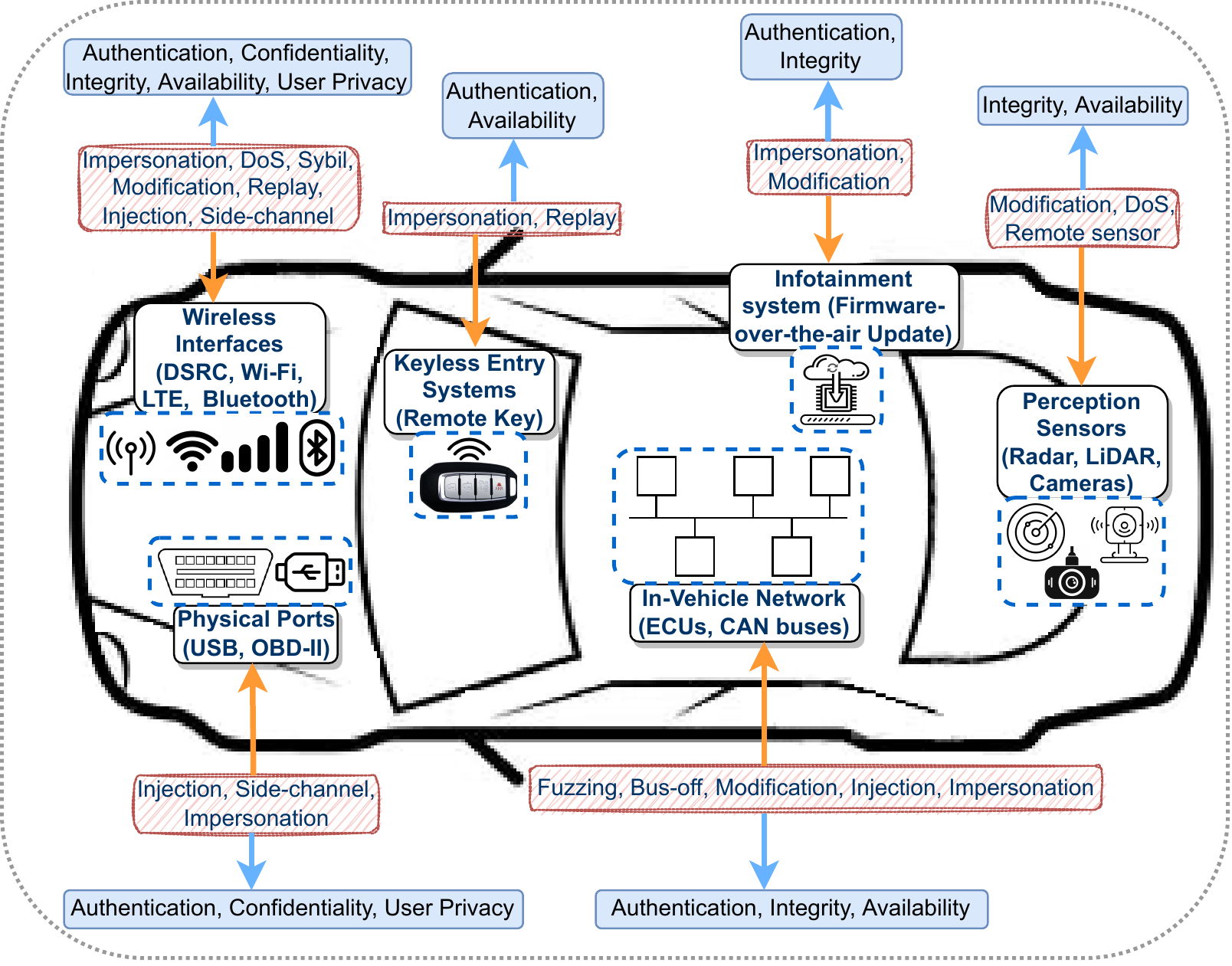}}
\caption{The Attack Surface Model for Connected and Automotive Driving System Operations.}
\label{AttackSurface}
\end{figure*}

Figure \ref{AttackSurface} displays CAV's potential attack surface model. Different CAV components (i.e., IVN, physical ports, wireless interfaces, keyless entry systems, infotainment system, and perception sensors) can be targeted to perform adversarial activities. The attackers can launch various attacks (e.g., impersonation, Sybil, replay, injection, bus-off, side-channel, and modification) on the automotive driving system to interrupt vehicle operations, get control of CAVs, and steal personal information. Essential security and privacy requirements are thus not satisfied while offering more comfort and intelligent vehicle services, leading to the insecure development of CAVs.

\subsection{Performance Measurement Parameters in ADPS}
It is vital to develop ADPS approaches that can detect abnormal events and seamlessly identify the sources of such situations in real-time. Since designed approaches are used in CAVs for robust results, it is important to measure the performance efficiency of the developed approaches to find anomalies. We describe significant performance parameters as follows, understanding the effectiveness of various ADPS approaches. It is considered that each occurrence $O$ that belongs to normal/regular events is considered as the positive sample, and other occurrences are considered as the negative samples \cite{VanWyk2019ITS}, \cite{Derhab2021ITS}. We have used notations (for Equations \ref{Accuracyformula}, \ref{Sensitivityformula}, \ref{Precisionformula}, and \ref{F1Scoreformula}) as $TP_{O}=$ the number of correctly identified positive samples, $TN_{O}=$ the number of correctly identified negative samples, $FP_{O}=$ the number of wrongly identified positive samples, and $FN_{O}=$ the number of wrongly identified negative samples.

\subsubsection{Accuracy}
It is measured based on the average of faultless predictions for abnormal and normal events or accurate and erroneous values (occurred in the system) from the total number of occurrences. The accuracy performance parameter formula is shown in Equation \ref{Accuracyformula}.
\begin{equation}\label{Accuracyformula}
Accuracy = \frac{\sum_{O}^{}\frac{TP_{O}+TN_{O}}{TP_{O}+FP_{O}+FN_{O}+TN_{O}}}{Number~of~Occurrences}
\end{equation}

\subsubsection{Sensitivity}
It evaluates the ratio of correctly found abnormal events or erroneous values from the number of the same anomalous/incorrect occurrences. Sensitivity is computed as per Equation \ref{Sensitivityformula}.
\begin{equation}\label{Sensitivityformula}
Sensitivity = \frac{\sum_{O}^{}\frac{TP_{O}}{TP_{O}+FN_{O}}}{Number~of~Occurrences}
\end{equation}

\subsubsection{Precision}
The proportion of erroneous values or abnormal events (among the forecasted anomalous/erroneous) from the number of the same anomalous/incorrect occurrences is called precision, and it is calculated based on Equation \ref{Precisionformula}.
\begin{equation}\label{Precisionformula}
Precision = \frac{\sum_{O}^{}\frac{TP_{O}}{TP_{O}+FP_{O}}}{Number~of~Occurrences}
\end{equation}

\subsubsection{F1 Score}
It is the harmonic mean of precision and sensitivity, and it can be calculated through Equation \ref{F1Scoreformula}.
\begin{equation}\label{F1Scoreformula}
F1~Score = 2 * \frac{Precision * Sensitivity}{Precision + Sensitivity}
\end{equation}

\subsubsection{Specificity}
It is the proportion of the number of correctly found injected packets (represented as $TN$) to the total number of actually injected packets (represented as $TN + FP$), and its formula is as written in Equation \ref{Specificity}, where $TN=$ true negative and $FP=$ false positive.
\begin{equation}\label{Specificity}
Specificity = \frac{TN}{TN + FP}
\end{equation}

\section{Articles Selection Methodology}
We first describe the considered article collection approach to identify related research papers for this survey article. After that, we discuss our results on a keyword searching process (that is carried out to include the most relevant papers for a more comprehensive and precise survey) and then explain how different research papers have been chosen for clear discussions.

\subsection{Article Collection Approach}
To collect relevant papers for the survey scope, we first selected precise keywords that are appropriate for ADPSs in connected and autonomous vehicles. Based on these chosen keywords, we then started keyword searching for the timeline of 2011-2021 on the topmost relevant scientific publication venues. The selected keywords for searching are \textit{Autonomous Vehicles/Cars}, \textit{Connected Vehicles/Cars}, \textit{Controller Area Networks}, \textit{In-Vehicle Networks}, \textit{Automotive Networks} as domain keywords, whereas \textit{Intrusion Detection} is taken into the account as a method keyword. Domain keywords mean the set of networks/services, which are based on the application areas. If a solution-based approach, technique, or mechanism is proposed/introduced toward the specific problem, it is considered a method keyword. We considered the following scientific publication venues to search for relevant papers. Conference proceedings\textbf{:} \textit{IEEE Symposium on Security and Privacy}; \textit{Network and Distributed System Security Symposium}; \textit{USENIX Security Symposium}; \textit{ACM Conference on Computer and Communications Security}; \textit{Annual Computer Security Applications Conference}; \textit{ACM ASIA Conference on Computer and Communications Security}; \textit{European Symposium on Research in Computer Security}; \textit{International Symposium on Research in Attacks, Intrusions and Defenses}; and \textit{IEEE European Symposium on Security and Privacy}; Journals\textbf{:} \textit{IEEE Transactions on Information Forensics and Security}; \textit{IEEE Transactions on Dependable and Secure Computing}; \textit{ACM Transactions on Privacy and Security}; \textit{IEEE Transactions on Intelligent Transportation Systems}; \textit{IEEE Transactions on Vehicular Technology}; \textit{Elsevier Computers and Security}; \textit{ACM Computing Surveys}; \textit{IEEE Communications Surveys \& Tutorials}; and \textit{IEEE Access}; We explicitly elucidate the method for selecting relevant papers and their results in the next section.

\subsection{Article Selection Method and its Results}
We follow certain criteria to include papers for more discussion in this survey article, and they are as follows:

\begin{itemize}
    \item A paper is included if it introduces/discusses the general concept of ADPS categories is included in it.
    \item A paper that proposes an ADPS approach, technique, or mechanism for CAVs.
    \item A set of ADPS performance measurements for CAVs are suggested/introduced.
    \item We have excluded poster/work-in-progress/demo papers in the process of relevant papers collection.
\end{itemize}

We performed keyword searching for the selected keywords, and the results are shown in Table \ref{QueryResults}. While considering the above-stated criteria, all keyword hits resulted in 3295 papers from the chosen scientific publication venues. We then studied all these papers based on their title/content to find relevant papers to the survey scope, resulting in 519 papers. Finally, we did an in-depth study of these articles to select papers for more discussions, and we found the most appropriate 75 papers for ADPSs for connected and autonomous vehicles. Out of these 75 papers, various ADPS approaches/techniques/mechanisms are proposed in 49 papers using different ADPS categories. We also did a query for ``Connected and Autonomous Vehicles" and ``Attack Detection and Prevention System" into the Web of Science (WoS) database\footnote{https://www.webofscience.com/wos}, and it resulted in 2785 and 1500 articles, respectively. In \cite{Liu2020ITS}, \cite{Alvarez2014ITS}, \cite{Jiang2019ICICS}, the consideration of road context is taken into account to improve the AV system efficiency. Here, the road context includes the road conditions (i.e., bend/joint/fork of roads, traffic light, and bump), nearby vehicles, pedestrians, weather conditions (i.e., fog, rain, and snow), lights conditions (i.e., the sunrise, sunset, and tunnel lights). Out of 75 papers are survey articles and other relevant papers in which the authors have discussed the ADPS for CAVs. Table \ref{QueryResults} displays query results for each keyword and the number of selected papers eventually for this survey, making it more straightforward for better understanding.

\begin{table}[!h]
\centering
\caption{Keywords Query Results on Selected Scientific Publication Venues}
\label{QueryResults}
\scalebox{0.78}{		
\begin{tabular}{|l|c|P{0.7in}|c|}  \hline
\rowcolor{lightgray} \hspace{0.4in}\textbf{Keywords} & \textbf{First Hits} & \textbf{Paper Title/ Content} &  \textbf{ADPS for CAVs}  \\ \hline
Autonomous Vehicles/Cars & 1350 & 250 & \multirow{6}{*}{75}  \\ \cline{1-3} 
Connected Vehicles/Cars & 840 & 174 &  \\ \cline{1-3} 
Controller Area Networks & 165 & 34 &  \\ \cline{1-3} 
In-Vehicle Networks & 245 & 39 &  \\  \cline{1-3} 
Automotive Networks & 295 & 14 &  \\  \cline{1-3} 
Intrusion Detection & 400 & 83 &  \\  \hline 
\rowcolor{lightgray} Total & 3295 & 519 & 75 \\
\hline
\end{tabular}}
\end{table}

\begin{figure}[!h]
\centering
{\includegraphics[width=0.45\textwidth]{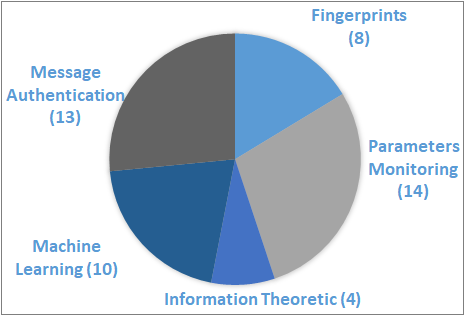}}
\caption{ADPS Categories Distribution with the Number of Papers in Selected Scientific Publication Venues}
\label{PieChart}
\end{figure}

Based on our literature study of various research articles, we have listed different ADPS categories, such as fingerprints, parameters monitoring, information-theoretic, machine learning, and message authentication. Attack detection and prevention solutions are mainly proposed based on these categories to find security threats and attacks in CAVs. After selecting 49 papers (that proposed ADPS approaches/techniques/mechanisms for CAVs), we have classified each paper under the specific ADPS category to understand their solution methodology to detect intrusions in CAVs. Figure \ref{PieChart} shows the number of papers for each ADPS category. These papers are considered for the literature in this survey for a detailed discussion, providing extensive information to the readers.

\section{Attacks Detection and Prevention Systems (ADPS) in CAVs}
An IDS is a software application or device that can find real-time incidents (performed by attackers to disrupt routine functionalities of the system) for any policy violations or suspicious actions by monitoring network traffic. An IDS can also act as a resilient protection technology for system security once standard technologies fail in the system \cite{GarciaTeodoro2009CS}. CAVs are enabled with many automated functionalities for a safe, more intelligent, and comfortable journey on the road. However, it is also essential to provide a high level of security in CAVs to avert infrastructure damages, human losses, and business crises and provide trustworthy services to the users. Thus, it is required to have an ADPS in CAVs that can offer effective identification and protection against attacks using either signature or anomaly-based solutions. 

\begin{figure}[!h]
\centering
{\includegraphics[width=0.48\textwidth]{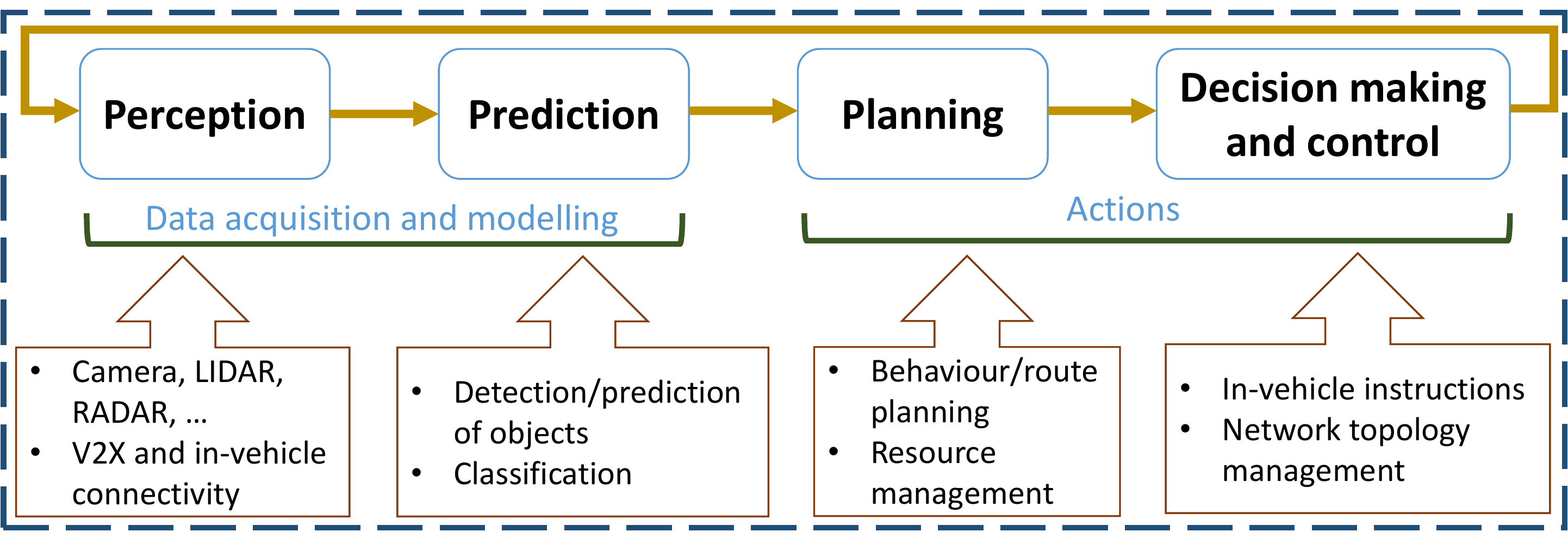}}
\caption{The Architecture Overview of Different Automotive System Implementation Stages in CAVs}
\label{SystemModelingCAVs}
\end{figure}

The automotive system architecture includes four stages to perform various activities in CAVs, i.e., (i) perception, (ii) prediction, (iii) planning, and (iv) decision making and control, as shown in Figure \ref{SystemModelingCAVs}. The first two stages are classified as data acquisition and modeling, whereas the third and the fourth stages are categorized as action parts. The first stage is to collect meaningful data from the camera, RADAR, LIDAR, and V2X/IVN connections. The second stage performs the detection, prediction, and classification of objects based on the given input data. The planning stage determines behavior and route direction for the classified data and manages the automotive system based on the available resources. The decision-making and control stage provides appropriate automotive instructions to the system to execute them adequately and manage the connected system components \cite{Jo2015AVCarITIE}. 

CAN is very limited in protecting various system components due to the unavailability of sufficient security features \cite{Aliwa2021CSUR}, \cite{Jo2021ITS}, \cite{Poudel2018TDSC}. CAVs are enabled with the outside world connectivity \cite{Woo2014ITS}, \cite{Petit2014ITS}, \cite{Humayed2017IoT} that allows adversaries for the execution of vulnerable activities in IVN remotely \cite{EASI2020NDSS}. Since CAVs have considerably less or null human intervention \cite{SAEAutomationLevels2021}, the failure of IVN operations may significantly impact the road infrastructures and people. Thereby, it is important to effectively identify possible security and privacy threats over the IVN that can reduce/avert damages in CAVs. Developing appropriate attack prevention solutions is also necessary to protect the automotive systems against such threats. A limited analysis has been presented on different types of ADPSs for CAVs and analyzed some of related research works in \cite{Kim2021CS}, \cite{Aliwa2021CSUR}, \cite{Jo2021ITS}, \cite{Wu2019ITS}, \cite{Lokman2019EJWCN}. We thus discuss various ADPS methods that are useful to detect intrusions in CAVs and to defend the automotive systems in such situations. Moreover, an extensive investigation is explicitly discussed on state-of-the-art research works for each ADPS category.

\subsection{Introduction and Analysis to ADPS Methods}
Attack detection and prevention solutions are proposed to detect security attacks to find vulnerabilities and protect the system from various attacks so that the security flaws are identified before they do real damage. In CAVs, there are mainly six types of ADPS categories, i.e., (i) fingerprints, (ii) parameters monitoring, (iii) information-theoretic, (iv) machine learning, (v) message authentication, and (vi) other approaches. We explicitly describe each ADPS category, as they are mainly used in designing attack detection and prevention solutions for CAVs.

\subsubsection{Fingerprints}
A fingerprint is a group of specific and unique configuration information that can identify devices, just as human fingerprints uniquely identify people. Data analysis can be applied to datasets such as network traffic and device configuration to extract the devices' fingerprints. In general, device fingerprinting can be classified into active or passive techniques; active techniques send specially crafted packets to probe the device, while passive techniques monitor the network traffic to detect patterns in the network traffic. Fingerprint-based IDS performs at the physical layer of the CAN bus, taking advantage of differences in physical properties, such as manufacturing variations, cabling, and aging, which allow ECU to be fingerprinted \cite{Wu2019ITS}. The digital fingerprints of the ECUs are then used to uniquely identify the sender of the CAN message. When the IDS detects an anomaly between the observed fingerprint of a CAN message and the profiled fingerprint of the sender's ECU, an alert is raised, and unauthorized or unknown nodes will be flagged.

As the characteristics of CAN signals are hardware-defined, the impersonation of CAN signals is difficult to tamper ECUs for an attacker without physical access. However, fingerprint-based IDS is ineffective against masquerade attacks \cite{Bhatia2021NDSS}, but this problem can be solved using a behavior-based IDS by analyzing the network traffic to create a signature. Attackers can compromise ECUs and use them to send malicious messages with the same physical fingerprint and remains undetected by fingerprint-based IDS. Although physical properties make excellent fingerprints, they vary with time due to changing environmental factors, especially the temperature and equipment aging of equipment \cite{Choi2018TIFS}. It will reduce the model's accuracy, which means that the IDS needs to be constantly updated with the latest fingerprints via periodic model retraining. Also, fingerprint-based IDS has a high computational demand due to the high sampling rates required to achieve accurate identification of devices by fingerprinting \cite{Kneib2018CCS}.

Driver style and behavior are affected by individual experiences and habits. In contrast, CAVs' driving behavior depends on the road conditions and driving model and should be less varied, more consistent, and more stable. Therefore, device fingerprinting should better identify devices due to the regularity of CAN traffic patterns and higher device identification accuracy. We describe related fingerprint-based solutions as follows.

A Clock-based IDS (CIDS) \cite{Cho2016USENIX} is developed to find the intervals of recurring in-vehicle instructional messages, and this helps to estimate the clock skews of ECU transmitters for fingerprinting details of ECUs. Fingerprints are then used to construct a baseline of ECUs' clock behaviors using the Recursive Least Squares (RLS) algorithm. Based on this baseline, CIDS performs a cumulative sum analysis to detect masquerade, fabrication, and suspension attacks over the CAN protocol, enabling fast identification of IVN intrusions (at a low false-positive rate of 0.055\%) and not missing any anomalies.

An attacker identification scheme, Viden (Voltage-based attacker identification), is proposed in \cite{Cho2017CCS} that finds the adversary ECU in the IVN based on the measurement and utilization of the voltage. Viden first determines the genuineness of the measured voltage signals during the ACK learning phase by checking whether the origin (of these signals) is from the legal message transmitter or not. The transmitter ECUs' voltage profiles are then updated as fingerprints based on the construction of the voltage measurements. Finally, an adversary ECU is detected in the IVN using the voltage profiles of an ECU. Based on the shown results on two actual vehicles and a CAN bus prototype, it is feasible to fingerprint ECUs through voltage measurements by Viden, thereby achieving a low false identification rate (of 0.2\%) to detect the adversary ECU in the system.

VoltageIDS is proposed in \cite{Choi2018TIFS} that aims to secure in-vehicle CAN networks through unique characteristics of CAN signals as fingerprints of ECUs. Taking masquerade and bus-off attacks for IVN into account, VoltageIDS is designed by observing two ECUs (one legitimate and another malicious) based on the sent identical signals to recognize the electrical characteristics of their messages, which is inherently challenging for the attackers to manipulate fingerprints. Further, VoltageIDS can also distinguish between a bus-off attack and errors in the system. The elevation of VoltageIDS is performed through actual vehicles and a CAN bus prototype setup that confirms the detection of intrusions in the in-vehicle CAN networks. 

Scission \cite{Kneib2018CCS} is proposed using fingerprint details (that can be extracted from CAN frames) to know the sender ECU's identification. Immutable physical characteristics from analog values are used to confirm the authorization of a sender ECU (to send evaluated messages), enabling to detect anomalies and the identification of compromised ECUs in the system. Scission's sender identification rate is 99.85 \% on average on two series production cars and a prototype setup. The results show that Scission can detect ECU-based attacks from compromised, unmonitored, and other added devices.

CAN is enabled with limited resources, and thereby, the high implementation costs or infringement of backward compatibility inhibits the deployment of CAN protocols in IVN to execute different functions properly. Thus, it has been found through an analysis in \cite{Foruhandeh2019ACSAC} that the state-of-the-art CAN ADPSs depend on multiple frames that are used to identify misbehavior of a certain ECU, but these frames are susceptible to a Hill-Climbing-style attack. Therefore, real-time intrusion detection and identification system, SIMPLE is developed to exploit the physical layer features of ECUs through a single frame, and ECUs can be effectively nullified. The results of the real-time vehicle and lab experiments with automotive-grade CAN transceivers show that the average equal error rates in SIMPLE are around 0.8985\% and 0\%.

The existing approaches offer good results to avert possible security challenges in CAN, but they require high computational effort and sampling rates. EASI \cite{EASI2020NDSS} is proposed by generating the fingerprint from a single symbol that improves the frame identification rate (of 99.98\%) with less computation effort. Further, it is demonstrated that comprehensive signal characteristics can be processed for voltage-based sender identification using machine learning algorithms. The results show that the computational requirements and the memory footprint are reduced by 142 and 168, respectively. Moreover, the classification problem is solved within 100 \textit{$\mu$s} with a training time of 2.61 seconds. 

The exposure of various real-world attack scenarios is designed to spoof the AV in \cite{Sun2021TIFS} so that it is possible to coerce the victim to make hazardous driving decisions, leading to a fatal crash. Based on the field experiments, the impacts of different attack scenarios are analyzed through a Lincoln MKZ-based AV testbed, and it confirms the access feasibility of the victim AV that enables the attacker to compromise the security and safety of the AV system. To address these challenges, challenge-response authentication and radio frequency fingerprinting schemes are developed to detect the above-discussed spoofing attacks, and the spoofing detection accuracy is achieved at a higher rate, 98.9\%.

A Voltage-based IDS (VIDS) effectively detects masquerade attacks that are launched based on a single attacker. The prior approaches can overcome single attacker-based masquerade attacks. However, a new voltage corruption strategy \cite{Bhatia2021NDSS} (based on a novel masquerade attack, named DUET) can be performed using two compromised ECUs to corrupt the bus voltages recorded by the VIDS: it is launched in a two-stage process (i) \textit{VIDS retraining mode}: manipulate a victim ECU's voltage fingerprint and (ii) \textit{VIDS operation mode}: impersonate the manipulated fingerprint. The execution of DUET shows the possibility of a novel masquerade attack in VIDS. To avert DUET in addition to other ECU masquerade attacks, a lightweight mitigation mechanism, RAndomized Identifier Defense (RAID) is proposed in \cite{Bhatia2021NDSS} using a unique protocol dialect (spoken by all ECUs on the CAN during the VIDS retraining mode). RAID is compatible with each ECU in frame format generation during VIDS retraining mode and protects against the corruption of ECUs' voltage fingerprints.

Table \ref{FingerprintsComparision} shows a comparative study of state-of-the-art fingerprints-based ADPS based on different attributes that give a better overview to understand the current scenario for attacks detection and protection using fingerprints.

\begin{table*}[!h]
\centering
\caption{Comparison of State-of-the-art Fingerprints-based ADPS for Different Attributes}
\label{FingerprintsComparision}
\scalebox{0.70}{		
\begin{tabular}{|p{1.1in}|c|p{2.2in}|P{1.3in}|c|c|p{2.1in}|}  \hline
\textbf{Scheme} & \textbf{Year} & \textbf{Objective(s)} & \textbf{Detection of Attacks} & \textbf{Impact on} & \textbf{False Positive} & \textbf{Limitations/Scope of Enhancement} \\ 
 & & & & \textbf{System/Device} & \textbf{Detection} & \\ \hline

Cho and Shin \cite{Cho2016USENIX} & 2016 & Find the intervals of recurring in-vehicle messages; & Masquerade, Fabrication, Suspension; & ECU; & Low & Ineffective during the injection of irregular messages; \\ \hline

Cho and Shin \cite{Cho2017CCS} & 2017 & Identify the attacker ECU; & Impersonation, Fabrication; & ECU; & Low & Ineffective for passive attacks by compromised ECUs;\\  \hline
Choi et al. \cite{Choi2018TIFS} & 2018 & Detect intrusions without involving any modifications in the system; & Bus-off, Masquerade; & ECU, CAN; & Low & Possibility of potential attacks due to the usage of frequent learning to understand power source condition variations;\\ \hline
Kneib et al. \cite{Kneib2018CCS} & 2018 & Identify the sender ECUs to assess its legality and detect attacks from additional and unmonitored devices; & Modification; &  ECU, CAN bus; & Low & Detection of compromised sender ECU for sent frames;\\ \hline
Foruhandeh et al. \cite{Foruhandeh2019ACSAC} & 2019 & Attack detection through a single frame with low computational and data acquisition costs; & Impersonation;  & ECU; & Low & Analysis of other crucial security threats;\\ \hline
Kneib et al. \cite{EASI2020NDSS} & 2020 & Determine attacks (based on compromised ECUs) with less computing resources; & Impersonation; & ECU; & Low & Management of corrupt electrical signals, Detect attacks from malicious ECU only; \\ \hline
Sun et al. \cite{Sun2021TIFS} & 2021 & Identify security vulnerabilities and enhance the
security and reliability in CAVs; & Replay, Impersonation; & Sensors; & --- & Management of corrupt electrical signals;\\  \hline

Bhatia et al. \cite{Bhatia2021NDSS} & 2021 & Propose a novel masquerade attack and identify the source of messages; & Masquerade, Modification, Replay; & ECU; & Low & Identification of benign and malicious bimodal distributions of voltage samples; \\ \hline
\end{tabular}}
\end{table*}

\subsubsection{Parameters Monitoring}
Parameters Monitoring-Based ADPSs detect attacks by monitoring parameters at the network and message levels in the IVN. It is a two-step process: First, baseline traffic is established to learn how the system behaves based on the parameters and to understand the regular traffic. Monitored traffic is then compared against the baseline, and the IDS flags for any abnormal traffic using anomaly-based detection. Some potential network-based detection sensors presented in \cite{Muter2010structured} are frequency, formality, location, range, correlation, protocol, plausibility, and consistency. Among the sensors, frequency is commonly used because most ECUs broadcast CAN frames regularly, and their transmission intervals can be easily observed \cite{Bozdal2021}.

Frequency-based IDSs are simple to apply and easy to analyze as an intrusion will disrupt the regularity of the CAN network and the frequency of the system \cite{Bozdal2021}. Besides, Parameters Monitoring-based IDSs have low computational requirements as they monitor parameters for abnormal flow or irregular traffic in the real-time network. However, an IDS that uses frequency as a parameter relies on the cyclic nature of CAN messages and is ineffective against non-periodic communications such as the locking and unlocking of door \cite{Bozdal2021}. In addition, the timing information of CAN traffic depends on the priority scheme of CAN, which may significantly change and affect the accuracy of the IDS \cite{Olufowobi2019Anomaly}. Lastly, fingerprint-based IDSs and parameters monitoring-based IDSs are vulnerable to masquerade attacks. The driving style of CAVs is determined by self-driving models and produces a standard network traffic pattern compared to human-monitored vehicles. Disruptions to the regularity of the CAV's CAN network will have a noticeable change from the baseline traffic, and it can be easily detected by parameters monitoring-based IDSs. We discuss relevant parameters monitoring solutions as follows.

The proliferation of ECUs and a wireless connectivity feature in present-day vehicles enable different functions and services, but it also opens the possibility of different security threats in CAN. In \cite{Cho2016CCS}, the bus-off attack, a new type of DoS, is proposed over the de facto standard IVN protocol, which exploits the error-handling scheme of IVNs aiming to shut down or disconnect uncompromised ECUs. The execution of a bus-off attack over actual IVN traffic on a CAN bus prototype and two real vehicles shows that this attack can be launched with the objectives of making uncompromised ECUs into defective ECUs and/or cessation of the complete automotive network. To address this challenge in IVN, a new defense mechanism is designed with two countermeasures as (i) \textit{indication of a bus-off attack}: look for consecutive error frames with an active error flag, and (ii) \textit{confirmation of a bus-off attack}: successful transmission of another message with the same ID. Another countermeasure can also be considered consecutive errors at the same bit position instead of frames.

IVN is enabled with many ECUs for various functions with Internet connectivity, and thereby, it has become a top-priority target point to launch automotive network system attacks. Thus, it is required to have compatible network mapping tools to report present security weaknesses and strengths of automotive networks. An automotive network mapping tool is developed in \cite{Kulandaivel2019USENIX} that supports finding vehicle ECUs and their communications with each other. However, there is a significant challenge in CAN, as CAN messages do not include the sender's information. Therefore, an automotive network mapper tool, CANvas, is designed to know the information of sender ECUs based on a pairwise clock offset tracking algorithm and finds the receiver ECUs using a forced ECU isolation technique. The results confirm that CANvas can precisely identify ECUs in the network and the senders and receivers of CAN messages on the open-source Arduino Due microcontroller.

A Dynamic Identifier Virtualization (VID) mechanism is developed in \cite{Sun2019Access} using random number sharing and substitution table methods to avert the analysis of CAN logs. Thus, generating valid messages by the adversary becomes more difficult. Thus, it reduces the possibility of spurious messages over the CAN bus. Implementing VID on real-time vehicles provides better results and identifies the adversary (attempting reverse engineering) through imposed time constraints.

Attackers should know the CAN message format to carry out suspicious activities in IVN, but this format is owned by Original Equipment Manufacturers (OEMs) and cannot be uniform even in different models of the same vehicle manufacture. Thereby, it is required to manually reverse-engineer the message format of each target vehicle, leading to inappropriate and time-taking procedures. A tool, LibreCAN \cite{Pese2019CCS} is developed that automatically translates most CAN messages with the least effort for reverse-engineering a complete CAN communication matrix for any vehicle. LibreCAN is designed with a three-phase procedure in which the first and second phases use two algorithms (i) signal extraction and alignment and (ii) defining the cut-off point for keeping pertinent signals with a high correlation value. The third phase is executed for snippeting recorded CAN data while performing body-related events. The achieved results through the third phase are highly accurate, and the second phase outcomes relatively outperform. In \cite{Pese2019CCS}, they also discussed recent steps taken to avert such attacks in IVNs. 

An attacker manipulates the transmission time of messages, aiming to spoof CAN messages by adding delays and thereby averting attack detection while launching cloaking attacks on the CAN bus. To combat this new type of masquerade attack, the execution of a cloaking attack is analyzed, and it is then systematically modeled to understand its success probability on the State-Of-The-Art (SOTA) and Network Time Protocol (NTP) IDSs \cite{Ying2019TIFS}. The evaluation of testbed setup and in an actual vehicle (i.e., UW EcoCAR) shows that the NTP-based IDS is comparatively effectual than the SOTA IDS in detecting masquerade attacks, and the cloaking attack is successful in NTP and SOTA IDSs. Experimental results on the collected data from UW EcoCAR verify that the average area deviation error (ADE) is 3.0\% for SOTA IDS and 5.7\% for NTP-based IDS.

The evaluation on an actual vehicle is performed in \cite{Othmane2020TDSC} for understanding the capability of the pearson correlation (due to popularity for data exploration) and unsupervised learning techniques, i.e., k-means clustering (as they do not need extended time for the implementation of attack detection mechanisms and may not rely on the context of the data.) as well as hidden Markov model (commonly used for better results). Vehicle's speed and Revolutions Per Minute (RPM) are mainly considered as reading parameters in \cite{Othmane2020TDSC} due to easy observation and safe injection of bogus speed/RPM reading messages on the CAN bus. 

The possibility of physical and cyber attacks is highly increased in IVN due to not having security features in wireless connectivity enabled CAN. To tackle these problems, a mechanism is first designed to extract real-time model values by observing the behavior of CAN bus messages. A specification-based automotive IDS based on CAN timing, SAIDuCANT \cite{Olufowobi2020TVT} is then developed using anomaly-based supervised learning techniques with the real-time model. Two new metrics, time to detection and false positives before the attack, are introduced to measure the performance of an IDS in terms of timeliness and classifier accuracy. Real-time vehicle implementation results of SAIDuCANT confirm the effective detection of data injection attacks with a low false-positive rate.

Human interaction modules are installed in CAVs for different functions, e.g., vehicle voice control systems, but the Automatic Speech Recognition (ASR) module may not detect accurate/correct voice commands or may proceed further through forged voice inputs, thereby leading to unexpected consequences. It is a considerable challenge to protect ASR systems from adversarial voice inputs in a hostile driving environment for driverless vehicles \cite{Wang2020RAID}. To address this problem, a three steps-based secure in-vehicle ASR mechanism, SIEVE \cite{Wang2020RAID} is developed that effectually identifies voice inputs given by the driver, passengers, or electronic speakers. SIEVE first does filtering of voice commands to distinguish the case of receiving the same signal multiple times in a short period from various sources, and it is done through autocorrelation analysis to find out the overlap of signals. In the second step, SIEVE checks whether a single-source voice input is from electronic speakers or humans based on a dual-domain identification technique through frequency domain-based acoustic characteristic, i.e., low-frequency energy attenuation. However, adversaries may attempt modulated voice inputs to disturb the ASR module. SIEVE uses time-domain parameters to detect non-human voice inputs effectively to detect modulated voice commands. The third step differentiates voice inputs whether given by the driver or the passengers, as it is required to prioritize the driver's voice command over the passengers for the smooth moving of a car. For this, SIEVE is developed by leveraging the directions of voice sources by calculating the time difference of arrivals on a pair of close-coupled microphones. Also, a spectrum-based detection technique is developed for better voice distinction between the driver and passengers. 

Localization of spiteful nodes during the node replacement/installation process is a remarkable challenge in CAN-based communication mechanisms, and the existing schemes are vulnerable to this issue \cite{Murvay2020Access}. New intrusion detection and localization system, TIDAL-CAN \cite{Murvay2020Access} is proposed by monitoring the propagation time of physical signals in which the time differences during signal propagation are calculated from the transmission point to the bus end. Furthermore, this variance is used as a location-based characteristic of the sender node to find malicious node installation/replacement and compromised nodes. The implementation results are mainly measured on the testbed setup by taking differential propagation delays into account, and they confirm that TIDAL-CAN can perform correct node classification without false positives even in attacks execution by compromised nodes. TIDAL-CAN can also identify transmitter nodes based on the attack method.

Conventional IDS methods are designed using time and frequency threshold values, and thereby they may result in higher false alert rates \cite{Bozdal2021}. A wavelet-based IDS, WINDS \cite{Bozdal2021} is designed through continuous wavelet transform to get the exact location of frequency components over the time axis, leveraging to first detect anomalies on the CAN bus. The analysis is then performed based on the scale domain to capture long-time and immediate short-time duration attacks. WINDS was evaluated on two datasets (generated through three commercial vehicles). The implementation results confirm that WINDS can reasonably achieve the attack detection rate even if an attack is immediately launched on the system.

Timing parameters of CAN frames can be used to create a secure channel that satisfies authentication, directly averting the requirement of cryptographic mechanisms in resource-constrained IVN for data transmission. However, this way can achieve a limited security level; thus, an adversary can launch different attacks on the CAN bus. In \cite{Groza2021TIFS}, an improved protocol is proposed through optimization algorithms (binary symmetric, randomized, greedy, and greatest common divisor) to schedule CAN frames cyclically and establish a covert channel for CAN traffic. Moreover, the proposed protocol can achieve higher data rates relatively on the covert channel due to the optimization of CAN traffic, enabling a 24-bit security level with six frames. The effective results can be achieved based on the proposed algorithms, i.e., a minimum inter-frame distance of 500 $\mu$s and an expected arrival time in the range of $\pm$ 5 $\mu$s.

When a CAN identifier (ID) sequence is configured through the IDs of CAN signals based on their order of occurrence, it will have a definite pattern. However, it is hard to identify the change in the corresponding pattern with a minimal number of attack IDs in a CAN ID sequence. In such cases, conventional IDSs are not effective. In \cite{Nam2021Access}, an IDS is developed using two bidirectional Generative Pre-trained Transformer (GPT) networks that allow using past and future CAN IDs. To reduce the Negative Log-Likelihood (NLL) value of the bidirectional GPT network, the proposed mechanism was inculcated for a typical ID sequence that detects an intrusion when the NLL value for a CAN ID sequence is larger than a pre-specified threshold.

Determining spoofing messages is a significant challenge due to the lack of sender identification and authentication in CAN. Thus, a delay-time-based technique, Divider, is previously proposed to find the sender ECU over the CAN bus. However, it is an ineffective solution while having ECUs with similar variations due to coarse time-resolution in Divider's measurement clock, making it challenging to distinguish ECUs. Moreover, another problem is the adaptability of a delay-time drift, caused by the temperature drift at the ambient buses \cite{Ohira2021AsiaCCS}. To deal with these challenges, a sender identification mechanism, PLI-TDC \cite{Ohira2021AsiaCCS} is developed using a super fine delay-time based Physical-Layer Identification (PLI) with Time-to-Digital Converter (TDC). PLI-TDC accurately identifies launched attacks on unmonitored and compromised ECUs. An accuracy rate of PLI-TDC is effective on a CAN bus prototype (of 99.67\%) and in a real vehicle (of 97.04\%), whereas a mean accuracy can be achieved around 99\% in PLI-TDC.

While considering the number of transferred messages and the importance of on-time message delivery in IVN, CAN-FD is better to satisfy high bandwidth and low latency requirements. However, CAN-FD is susceptible to masquerade attacks due to the unavailability of authentication protocols and adequate defense measures. In \cite{Xie2021TVT}, a dual-pointer solution, forward-backward exploration is proposed based on three methods, i.e., combination enumeration, forward exploration, and backward exploration for secure transfer of independent CAN-FD messages in IVN. In this solution, the Message Authentication Code (MAC) size of each message is dynamically balanced through dual-pointer movement rules until the total payload no longer increases, providing enhanced security by increasing the total MAC size of CAN messages, and the forward-backward exploration achieves better time efficiency by completing the exploration process. Thereby, this solution can be applied for trustworthy CAN-FD message transmission in IVN.

Table \ref{ParametersMonitoringComparision} displays a comparison outline of state-of-the-art parameters monitoring-based ADPS based on different attributes, making it easier to understand the security severity in CAVs through parameters detail.

\begin{table*}[!h]
\centering
\caption{Comparison of State-of-the-art Parameters Monitoring-based Systems in Different Attributes}
\label{ParametersMonitoringComparision}
\scalebox{0.70}{		
\begin{tabular}{|p{1.1in}|c|p{2.3in}|P{1.05in}|P{0.85in}|P{0.8in}|p{2.1in}|}  \hline
\textbf{Scheme} & \textbf{Year} & \textbf{Objective(s)} & \textbf{Detection of Attacks} & \textbf{Impact on System/Device} & \textbf{False Positive Detection} & \textbf{Limitations/Scope of Enhancement} \\ \hline

Cho et al. \cite{Cho2016CCS} & 2016 & Expose possible vulnerabilities based on a bus-off attack in IVN and propose a scheme to detect and protect this attack; & Bus-off, Injection; & ECU, CAN bus; & Low & Detection of other important security threats;\\ \hline

Kulandaivel et al. \cite{Kulandaivel2019USENIX} & 2019 & Develop a network mapping tool to identify the senders and receivers of messages; & Modification, Bus-off, Fabrication; & ECU; & Low & Identification of itentions of transferred messages; \\ \hline

Sun et al. \cite{Sun2019Access} & 2019 & Make CAN logs complex to avoid generation of valid messages illegally; & Injection, Replay;  & CAN bus; & --- & Requirement of high computational resources, Less complexity in the data field of messages to ;\\ \hline

Pese et al. \cite{Pese2019CCS} & 2019 & Develop a message format translation tool of most CAN messages with minimal effort to exploit vulnerabilities faster; & \hspace{0.5in}--- & CAN messages; & Average & Require involvement of vehicle OEMs to mutually agree on specific attributes; \\ \hline

Ying et al. \cite{Ying2019TIFS} & 2019 & Detect masquerade and time-based attacks; & Replay, Masquerade; & CAN; & Low & More noise in the IDS for messages; \\ \hline

Othmane et al. \cite{Othmane2020TDSC} & 2020 & Identification of injection attacks in the given vehicle status, i.e., "under attack" or "no attack"; & Injection, Modification; & CAN bus; & Average & Not conclusive results for in-motion vehicle messages; \\ \hline

Olufowobi et al. \cite{Olufowobi2020TVT} & 2020 & Detect threats using real-time schedulability response time analysis; & Injection, Impersonation, Replay; & ECU, CAN bus; & Low & Ineffective in attacks classification;\\ \hline

Wang et al. \cite{Wang2020RAID} & 2020 & Detect vulnerabilities in the ASR module and provide protection against such threats; & Impersonation, Replay; & ASR module; & Low & Better results in different scenarios, i.e., input methods, accuracy in noisy environment, and overlapping input commands; \\ \hline

Murvay et al. \cite{Murvay2020Access} & 2020 & Estimation the relative location of a transmitter node on the CAN bus; & Bus-off, Replay, Modification; & CAN; & Low in specific conditions & Better results in different circumstances; \\ \hline

Bozdal et al. \cite{Bozdal2021} & 2021 & Identify the behavior change location in the CAN traffic; & Impersonation, Replay, Bus-off; & CAN bus; & Low in certain conditions & Effective when message frequency is used in the attack scenario; \\ \hline

Groza et al. \cite{Groza2021TIFS} & 2021 & Design efficient attacks detection and authentication scheme using optimization algorithms; & Replay, Bus-off; & ECU, CAN bus; & Low  & Detection of other important security threats;\\ \hline

Nam et al. \cite{Nam2021Access} & 2021 & Detect security threats using past and future CAN IDs for better attack pattern identification; & Replay, Injection; & CAN bus; & Low & Ineffective to detect manipulated messages, sent by compromised ECUs;\\ \hline

Ohira et al. \cite{Ohira2021AsiaCCS} & 2021 & Improve the identification accuracy of message senders using super fine delay-time with time-to-digital converter method; & Impersonation; & ECU; & Low & Detection of other important security threats;\\ \hline

Xie at al. \cite{Xie2021TVT} & 2021 & Detection of masquerade attacks in real-time transmitted CAN-FD messages; & Masquerade; & CAN bus; & --- & Need more computational resources; \\ \hline

\end{tabular}}
\end{table*}

\subsubsection{Information-Theoretic}
Information theory is the mathematical treatment of the concepts, parameters, and rules governing the transmission of messages through communication systems. Entropy is a crucial measure in information theory, which relates to the measure of disorder and the uncertainty associated with a random variable. In computer networks, IDS has applied entropy to detect threats based on anomalous patterns in the network. Entropy-based anomaly detection algorithms characterize the expected behavior of a set of data based on their level of statistical entropy \cite{Cover1991Elements}. The two key underlying assumptions of entropy-based anomaly detection are that the entropy of messages generated by the information source exhibits stable statistical characteristics and the anomalies introduce significant deviations in the statistical characteristics of the entropy. Traffic in IVN is mainly cyclic, and the information entropy is low and stable \cite{Marchetti2016Evaluation}, \cite{Wang2017Hardware}, making entropy-based anomaly detection suitable. 

Since Information-Theoretic-based IDSs depend on the data information and flow, they are independent of CAN messages' content. Hence, it can be applied to any traffic, even proprietary messages. However, they are ineffective against attacks that target the content of CAN messages, i.e., masquerade attacks. The main limitation of Information-Theoretic-based IDSs that it is ineffective against low-volume attacks, in which the attackers inject only a few packets per second and avoid increasing the entropy of the system \cite{Marchetti2016Evaluation}. Entropy-based IDS is ineffective against CAN messages with high entropy even during normal operations. Due to CAVs' consistent driving style, the IVN traffic of CAVs should have lower randomness and higher entropy stability than that of human intervention vehicles. Disruptions to the IVN traffic's entropy should be more noticeable and significantly increase the system's entropy. We describe information-theoretic-Based solutions as follows.

To detect the feasibility of modification and replay attacks in CAN, an IDS solution in \cite{Groza2018TIFS} is designed using Bloom filters (considering its efficient time memory trade-off) that verifies frame periodicity through message identifiers and contents of the data field. Thus, it effectively detects modified frames by testing the frame's content. In contrast, duplicate frames are identified through an IDS even an attacker attempts to replay frames in the optimal time frame. This work mainly shows the possibility of using Bloom filters in developing CAN-based IDSs to achieve better results in detecting intrusions in the system.

The issues of random cable connectivity for a short duration and the Intermittent Connection (IC) fault are directly linked to the system performance. Therefore, the possibility of system-level failures and system performance degradation can be increased if these problems are not addressed effectively in CAN. Thus, it is essential to precisely detect and localize the IC fault for better health management of CAN-based network systems. To address this problem, a systematic and practical IC fault diagnosis framework \cite{Zhang2019TVT} is developed for CAN-based on the collected error event pairs from the data link layer. The scheme extracts the positive and negative information from these error event pairs to combine them for diagnosing the IC faults. The results of the proposed framework in \cite{Zhang2019TVT} can be used as insights into the characteristics of IC faults for quick diagnosing during different circumstances that provide better system maintenance, improving the system reliability.

Considering only color/textural information of images is valuable for semantic reasoning. However, combining semantic information and depth information of images can substantially enhance scene parsing performance, especially in wrongly categorized based on only RGB features. Therefore, the Built-in Depth-Semantic Coupled Encoding (BDSCE) \cite{Liu2020ITS} module is proposed by integrating Red Green Blue (RGB) and depth features that present important depth-discriminative features selectively. The BDSCE is congruent with existing CNN-based mechanisms and can offer better scene parsing results to address misclassification. The Depth-Semantic Coupled Encoding Network (CEncNet) framework is developed using the BDSCE module to extend the conventional deep scene parsing. The implementation results on the datasets, Cityscape \cite{Cordts2016CVPR} and  KITTI Vision Benchmark Suite \cite{Geiger2013JBR}, confirm that CEncNet achieves better performance than the traditional mechanisms. The extensive experiments also show the effectuality of the BDSCE module for vehicle detection and road segmentation in city areas.

Extensibility plays a significant factor in the automotive network, as it is developed based on the Electrical/Electronic (E/E) architectures. However, this optimization objective should be extensively considered in the design of IVN for the implementation of new functionality or modification in the existing functionality. To consider this problem in IVN, a new extensibility model \cite{Xie2021ITSOptimizing} is developed for CAN using the Mixed-Integer Linear Programming (MILP) algorithm for mid-sized signal sets and the simulated annealing-based heuristic algorithm for industry-sized signal sets. Moreover, the corresponding extensibility metric for CAN-FD is designed. The results (extensive implementations through synthetic signal sets) show the effectiveness of the proposed approaches in \cite{Xie2021ITSOptimizing}.

A comparative description of state-of-the-art information theoretic-based ADPS is given in Table \ref{InformationTheoreticComparision} that makes it easier to understand the importance and effectiveness of using information-theoretic approaches in CAVs.

\begin{table*}[!h]
\centering
\caption{Comparison of State-of-the-art Information Theoretic-based Systems in Different Attributes}
\label{InformationTheoreticComparision}
\scalebox{0.70}{		
\begin{tabular}{|p{1.1in}|c|p{2.3in}|P{1.05in}|P{0.85in}|P{0.8in}|p{2.1in}|}  \hline
\textbf{Scheme} & \textbf{Year} & \textbf{Objective(s)} & \textbf{Detection of Attacks} & \textbf{Impact on System/Device} & \textbf{False Positive Detection} & \textbf{Limitations/Scope of Enhancement} \\ \hline

Groza and Murvay \cite{Groza2018TIFS} & 2018 & Effective detection of replay or modification attacks; & Replay, Modification; & CAN bus; & Average & Requirement of advanced security features for threats protection; \\ \hline
Zhang et al. \cite{Zhang2019TVT} & 2019 & Effective detection and accurate localization of the intermittent connection (IC) fault; & \hspace{0.5in}--- & CAN; & --- & Requirement of a better method to satisfy the objective in a complex network;\\ \hline
Liu et al. \cite{Liu2020ITS} & 2020 & Enhance scene parsing performance results especially for clear depth distinction and misclassified through RGB-only features; & \hspace{0.5in}--- & CAN; & --- & Advancement of performance results; \\ \hline
Xie et al. \cite{Xie2021ITSOptimizing} & 2021 & Provide signal packing in the context of extensibility for CAN-FD; & \hspace{0.5in}--- & Network bandwidth utilization; & --- & Improvement in performance results; \\ \hline

\end{tabular}}
\end{table*}

\subsubsection{Machine Learning}
Machine Learning (ML)-based IDSs have been deployed extensively in network security due to their ability to detect unknown attacks via anomaly detection through Artificial Intelligence (AI). The learning process starts by analyzing provided data set to identify patterns, learn automatically using mathematical models, and extract useful information to make better predictions. Machine learning can be classified mainly into supervised and unsupervised machine learning. Unsupervised learning algorithms can understand and model the typical profiles of the network and report anomalies without any labeled data set \cite{Aliwa2021CSUR}. On the other hand, supervised learning algorithms learn from labeled training data and predict future events based on the past.

Most machine learning-Based IVN IDSs can be classified into the machine learning techniques applied; traditional machine learning and deep learning. Traditional machine learning techniques, including Support Vector Machine (SVM), Decision Tree (DT), Random Forest (RF), and Multi-Layer Perceptron (MLP), can be applied to IVN IDSs for understanding the pattern of CAN network data to learn the expected behavior of the system \cite{Moulahi2021Access}. Deep learning techniques use artificial or deep neural networks, algorithms inspired by the human brain. It works by repeatedly learning, understanding, and tweaking the model to achieve the best outcome, similar to how a human would conclude. A multi-layered structure of algorithms is applied to identify patterns and classify different types of information. The individual layer of the neural networks acts as a filter that increases the likelihood of detecting and predicting a correct outcome \cite{Song2020vehicle}.

The main advantage of ML-based IVN IDSs is their strength in detecting unknown attacks by reviewing large volumes of data and discovering trends and patterns that would not be apparent to humans. Furthermore, the model continuously improves accuracy and efficiency as more data is fed into the model. Deep learning also avoids the complex feature extraction step compared to traditional ML. The main disadvantage of ML-based IVN IDSs is the high computational requirement compared to the previous categories of IDSs \cite{Moulahi2021Access}. In addition, a large data set is required to train the model and valuable data set is rare, especially those with attacks or abnormal traffic. The ensemble method is a technique that combines several base models in order to produce one optimal predictive model. It has been shown to achieve the desired accuracy and robustness \cite{Yang2019Access} and to overcome the limitation of machine learning techniques.

Compared to a typical vehicle, CAV relies on multiple sensors, including cameras, radars, and LIDARs. These CAV sensors and ECUs produce large quantities of highly relevant data for analysis with machine learning techniques. It also helps to enhance the accuracy and performance of the existing training models. In addition, data could be collected remotely for CAVs, which increases the ease of data collection and the volume of available training data. Related machine learning-based solutions are discussed as follows.

The number of CAVs will increase in the near future, and it is vital to detect abnormalities and discern their sources to provide a seamless experience for driverless vehicles in real-time. Therefore, anomaly detection and identification techniques are developed by effectively integrating a Convolutional Neural Network (CNN) and Kalman Filtering (KF) to find CAV systems' abnormal activities. CNN is first applied to time-series data (acquired from various sensors), and images are then generated from real-time raw sensor data to classify them as abnormal. After that, a general framework is proposed using CNN and KF with a $\chi^{2}$-detector (named CNN-KF) to detect anomalies in CAVs. The experimental results of proposed approaches (only CNN, only KF, and CNN-KF) are evaluated based on accuracy, sensitivity, precision, and F1 score. CNN-KF framework collectively outperforms in these performance parameters for anomaly detection and identification \cite{VanWyk2019ITS}.

For a compelling and comfortable journey, CAN is used in automotive systems (e.g., CAVs) to execute different functions without/less human interaction. However, such automated systems are vulnerable to known and unidentified security threats, so it is necessary to detect such incidents early to avert infrastructure damage and loss of human life on the road. A Dynamic Ensemble Selection System (DESS) \cite{Yang2019Access} is developed for anomaly detection in which the system includes two-class and one-class classifiers to identify fault types (from the training data set) and unknown fault types. Moreover, the network features are extracted from the physical-layer information, and the base classifiers are then trained based on these network features. The implementation was carried out on the data set, and the analysis confirms that anomaly detection robustness and adequate accuracy can be achieved through DESS with better results than other methods, even in different fault types.

Different types of sensors in modern vehicles collect data from a vehicle and nearby objects to provide meaningful information to the vehicular communication system, enabling it to make better decisions while moving. However, this collected data from sensors are susceptible to different inconsistencies (caused by errors, cyberattacks, and/or faults), and thereby, the direct usage of sensor-generated data may lead to accidents on the road \cite{Javed2020ITS}. A Multi-Stage Attention scheme with a Long Short-Term Memory-based CNN (MSALSTM-CNN) \cite{Javed2020ITS} is developed to detect anomalies from sensor-generated data, helping to avoid fatal casualties by CAVs. In MSALSTM-CNN, multi-source sensor readings are first classified as either ordinary or abnormal data, and it then concentrates on different values of streaming readings to understand their importance. A weight-adjusted fine-tuned ensemble, WAVED, is also proposed through the optimal weight vector of classifiers to set a unique voting weight to anticipate each classifier and identify anomalous actions. The experimental results demonstrate that the MSALSTM-CNN can achieve a better anomaly detection rate in the case of single and mixed anomaly types. Thus, fatal casualties (caused due to anomalous data) can be reduced through MSALSTM-CNN.

A communication network in CAVs is vulnerable due to the unavailability of security features in CAN and connectivity with the outside network for meaningful data exchanges, resulting in different types of suspicious activities. To deal with such a situation, a deep learning-based IDS is designed in \cite{Ashraf2020ITS} to find out malicious network activities from IVN, V2V, and V2I networks of autonomous vehicles. A Long-Short Term Memory (LSTM) autoencoder algorithm is developed using deep learning architecture to detect intrusive incidents from the gateways of AVs. On the UNSW-NB15 dataset \cite{Moustafa2015MilCIS}, the proposed IDS can achieve 98\% accuracy in detecting different types of attacks, whereas 99\% accuracy is achieved on the database of car hacking for in-vehicle communications.

IVN is susceptible to various network-based attacks due to the lack of security features in CAN and V2X connectivity with associated ECUs through the gateway ECU. Therefore, a CAN Bus message Attack Detection Framework (CAN-ADF) \cite{Tariq2020CAS} is proposed to generate abnormality, detect anomalies, and validate the system performance for the CAN bus architecture. A rule-based method is designed from different network traffic characteristics and Recurrent Neural Networks (RNNs) for anomaly detection. A large number of CAN packets are collected from different vehicles to analyze the performance of CAN-ADF, showing an average accuracy of 99.45\%. A visualization tool is designed to monitor the CAN bus traffic status, and it displays found attacks in the IVN system. CAN-ADF can be combined with other attack detection methods to identify a range of anomalies effectively. 
 
The automotive system should have a trustworthy environment for reliable communications, as information plays a significant role in CAVs. In \cite{Islam2020ITS}, a graph-based four-stage IDS is proposed to detect various attacks in CAN in which a graph-based technique first finds abnormal patterns in the dataset. After that, the median test and chi-squared methods are applied to differentiate the two data distributions. The experiments exhibit that the misclassification rate is comparatively low for the proposed IDS in \cite{Islam2020ITS}, i.e., 4.76\% for replay, 5.26\% for DoS, and  10\% for fuzzy attacks detection. All spoofing attacks can accurately be detected through the proposed method in \cite{Islam2020ITS}, and it can achieve better accuracy up to 13.73\%.

\begin{table*}[!h]
\centering
\caption{Comparison of State-of-the-art Machine Learning-based Systems in Various Features}
\label{MachineLearningComparision}
\scalebox{0.70}{		
\begin{tabular}{|p{1.1in}|c|p{2.3in}|p{1.1in}|P{0.85in}|P{0.8in}|p{2.1in}|}  \hline
\textbf{Scheme} & \textbf{Year} & \textbf{Objective(s)} & \textbf{Detection of Attacks} & \textbf{Impact on System/Device} & \textbf{False Positive Detection} & \textbf{Limitations/Scope of Enhancement} \\ \hline

VanWyk et al. \cite{VanWyk2019ITS} & 2019 & Detection of abnormalities and source identification of attackers; & Injection and impersonation attacks; & Sensors & Low & Provided results in certain conditions; \\ \hline
Yang et al. \cite{Yang2019Access} & 2019 & Identify abnormality in advance and develop an accurate and stable anomaly detector; & Delayed operations; & CAN communications; & Average & Detection of transient faults; \\ \hline
Javed et al. \cite{Javed2020ITS} & 2020 & On-time detection of anomalies in CAVs; & System damage; & Automotive system; & Low & Requirement of the prediction votes above 50\% \\ \hline
Ashraf et al. \cite{Ashraf2020ITS} & 2020 & Discover susceptible actions over IVN, V2V, and V2I networks; & DoS, replay, and impersonation attacks; & V2X and IVN connectivity; & Low & Require to improve an IDS for accurate attacks categorization; \\ \hline

Tariq et al. \cite{Tariq2020CAS} & 2020 & Detect CAN bus attacks; & Replay, injection, and bus-off attacks; & CAN bus; & Low & Able to detect specific attacks;\\ \hline
Islam et al. \cite{Islam2020ITS} & 2020 & Detect attacks in CAN; & Replay, impersonation, injection, and bus-off attacks; & CAN; & Low & Possibility to reduce the misclassification of attacks;\\ \hline
Moulahi et al. \cite{Moulahi2021Access} & 2021 & Comparative study of machine learning approaches for attacks detection; & Replay, injection, and impersonation attacks; & CAN; & --- & High computational resources and enough data; \\ \hline
Derhab et al. \cite{Derhab2021ITS} & 2021 & Find intrusions by assembling the CAN packets into windows to classify the traffic; & Impersonation, replay, and injection attacks; & CAN; & Low & More efficient and lightweight IDS; \\ \hline
Liu et al. \cite{Liu2021TDSC} & 2021 & Protect CAVs against perception error attacks; & Impersonation and injection attacks; & Sensors; & Low & Need to improve for optimization results; \\ \hline
Han et al. \cite{Han2021TIFS} & 2021 & Detection and identification of anomalies through the periodic event-triggered interval; & Modification, replay, injection, bus-off attacks; & ECU, CAN bus; & Low & Required to design a method for better intrusion detection time;\\ \hline

\end{tabular}}
\end{table*}

To deal with the problem of unavailability of sender information (in sent messages over the CAN bus), an IDS is developed by using various machine learning approaches, i.e., Support Vector Machine (SVM), Decision Tree (DT), Random Forest (RF), and Multi-Layer Perceptron (MLP) for CAN \cite{Moulahi2021Access}. The proposed IDS is applied to the KIA Soul car dataset to detect intrusions and the type of attacks based on a set of classifiers. The implementation results state that the RF classifier can achieve better results than DT, SVM, Recurrent Neural Network (RNN), Hierarchical Temporal Memory (HTM), and Hidden Markov Model (HMM) classifiers in the same context. Moreover, the precision result of SVM, MLP, RF, and DT is superior to HMM and RNN, but it is moderately poor than HTM.

In \cite{Derhab2021ITS}, a Histogram-based Intrusion Detection and Filtering (HIDF) mechanism is developed by combining a window-based IDS and filtering approach to identify intrusions based on windows and do the filtration of regular CAN packets from an attack window. An intrusion detection model is first developed using histograms of CAN traffic to understand a distinctive structure for different CAN traffic classes. Furthermore, a one-class SVM attack model is developed using regular CAN traffic and implemented with four attack variants, i.e., Gear, RPM, Fuzzy, and DoS. The experimental results based on two datasets demonstrate that the HIDF can accurately classify through a window, and the filtering system is capable of filtering out standard packets from abnormal windows with more than 95\% correctness. 

CAVs are enabled with multiple sensors to collect relevant data and use it as inputs in various vehicle driving decisions. Thus, it is vital to ensure the reliability of this sensory information for errorless execution of different operations in CAVs. A Perception Error Attack (PEA) can fail sensors to perceive the surrounding driving environment accurately, and thereby, captured data may be faulty, leading to unexpected consequences. To address this issue for AVs, a countermeasure approach is proposed, LIDAR and Image data Fusion for detecting perception Errors (LIFE) \cite{Liu2021TDSC} that identifies PEAs by evaluating the data consistency between LIDAR and camera image through object matching and corresponding point techniques. Thus, LIFE can detect various sensory data anomalies, i.e., LIDAR spoofing, camera blinding, false positives/negatives during object identification, LIDAR/camera rotation error, and LIDAR saturation/distance measurement error. Since anomalies are detected through LIFE, they can be forwarded to the driving system to make appropriate decisions. The evaluation results on the KITTI dataset show that LIFE provides average performance. However, LIFE can be improved for better performance results, i.e., reduce the number of false alarm instances for high intrusion detection efficiency and minimize the requirement of additional settings in existing autonomous vehicles.

To protect against maleficent packet attacks in CAVs, it is required to find anomalies effectively; otherwise, the automated system may lead to unexpected situations, resulting in risky commute and infrastructure damages. Thus, an event-triggered interval-based mechanism is proposed using machine learning to identify abnormalities and detect attacks in IVN \cite{Han2021TIFS}. Four attack scenarios are first defined based on CAN messages to understand normal and malicious driving data in the context of IVN. The event-triggered interval of CAN identities is then analyzed and measured in their statistical instants by considering the fixed time window. The results of the experiment over actual driving data demonstrate that the proposed method in \cite{Han2021TIFS} can quickly identify anomalies and achieve better performance in attack type identification, time, and anomaly detection.

Table \ref{MachineLearningComparision} shows an analogical study of state-of-the-art machine learning-based ADPS. It compares recent ADPS solutions to know their efficacy in various attributes.

\subsubsection{Message Authentication}
Message authentication is used widely in information security to ensure that data integrity and authenticity are preserved while in transit and allow the receiver to verify the source of messages. Common message authentication mechanisms include MACs, Authenticated Encryption (AE), and digital signatures. CAN does not have a built-in authentication process, making it vulnerable to masquerade attacks. However, the deployment of cryptographic methods is complex due to the CAN protocol's low throughput and limited bandwidth. Researchers have looked into several ways, including message authentication and covert channels, to meet the specific deployment criteria in IVNs. 

The most important benefit of message authentication is the protection against masquerade attacks, as CAN is a broadcast protocol without authentication. However, most message authentication solutions require modifications of the CAN protocol or the introduction of additional information on the CAN frame. In addition, generating MACs and checksums increases the computational workload of the already resource-constrained ECUs. As compared to IDSs, message authentication is harder to deploy on existing vehicles as it requires either the manipulation of the CAN hardware or the addition of new hardware such as key server and Trusted Platform Module (TPM). We discuss relevant message authentication mechanisms as follows.

CAN control messages are crucial in IVN, but the sender information is unavailable in CAN messages, leading to denial of service, impersonation, and data alteration challenges. A security protocol is proposed in \cite{Woo2014ITS} to deal with such challenges using authentication and data encryption mechanisms. The proposed scheme is designed with a MAC (to remedy the fixed data payload size of CAN data frames) and key management approach to provide secure exchanges between in-vehicle ECUs and external devices. The experimental results based on a manufactured ECU demonstrate the possibility of an attack over wireless connectivity through a malicious smartphone app. Performance analysis shows that the proposed protocol in \cite{Woo2014ITS} takes less computational resources, but it is susceptible to encryption key compromising, authentication attacks, and session key leakage.

A compromised Compact Disk (CD) player can execute crucial operations, i.e., accelerate in CAVs. A LIghtwEight Authentication scheme for CAN, LEIA \cite{Radu2016ESORICS} is proposed to verify ECUs and protect them from compromised vehicle components. LEIA runs under the exact time and bandwidth constraints of automotive applications, and it is designed using unidirectional authentication in which a method of signaling technique is applied with the session key to check whether any of the subscribed ECUs follows the synchronization/authentication process or not. Security analysis of LEIA confirms the protection against chosen-plaintext attacks. 

Present-day automobile systems are susceptible to various security threats, compromising vehicle travelers' physical safety. A new ECU architecture is proposed in \cite{Poudel2018TDSC} for automotive cyber-physical systems to satisfy security and performance attributes effectively. It is implemented on the Xilinx Automotive Spartan-6 field-programmable gate array and NXP iMX6Q SABRE automotive board. The results confirm lower computation time and response time in \cite{Poudel2018TDSC}.  

Sharing a secret key in CAN is a challenging task due to the broadcast nature of the CAN bus architecture. A protocol suite \cite{Groza2019Access} is suggested for the secure exchange of keys over the CAN bus, and it is a combination of time-triggered mini-max and randomized delay key negotiation, which allows piggybacking frames with the keys' portions for secure computation of a session key. Moreover, CAN frames can be sent through the Diffie-Hellman (DH) version of the Encrypted Key Exchange (EKE) and Simple Password Exponential Key Exchange (SPEKE) protocols. The implementation was carried out on high-end controllers over Infineon Aurix cores (i.e., TC297 and TC277), and the outcome achieves reasonable results based on simple bus-based key negotiation and EKE/SPEKE-DH key sharing approaches.

A keyless entry system is more convenient for CAV users, but it is susceptible to signal-relaying and network range attacks, making it difficult to distinguish an authorized door unlock request from a spiteful signal. An RF-fingerprinting technique, HOld the DOoR (HODOR) \cite{Joo2020NDSS} is proposed to identify attacks in the keyless entry systems. HODOR is developed as a sub-authentication mechanism based on ultra-high frequency band RF signals to implement on existing authentication processes (of keyless entry systems) without any modifications. The implementation results show that HODOR provides satisfactory results as the average false positive rate of 0.27\% and the false-negative rate of 0\% while considering the detection of simulated attacks. HODOR achieves the false-positive rate of 1.32\% to detect legal key determination under the non-line-of-sight conditions.

CAN communications are unprotected in IVN, leading vehicles towards adversarial activities based on wired/wireless attacks. An efficient authentication protocol suite is proposed in \cite{Palaniswamy2020TIFS} to provide a secure connection for transmitting remote frame requests and updating session keys between in-vehicle ECUs and external devices through entity authentication and key management using ECC. The proposed protocol in \cite{Palaniswamy2020TIFS} achieves better security and performance results than \cite{Woo2014ITS}, but it is vulnerable to encryption key compromising and authentication attacks.

Modern cars are configured with different ECUs, including safety-critical, and the possibility of remote access is demonstrated to perform malicious activities in the CAN, allowing an attacker to control a vehicle. The existing message authentication protocols for CAN are either vulnerable to masquerade attacks or require hardware modification to protect against such attacks. A new Mutual AUTHentication scheme, MAuth-CAN \cite{Jo2020TVT} is proposed using a unique session authentication key (computed through its seed value of an ECU) for each ECU to resist masquerade and bus-off attacks. The performance of MAuth-CAN was evaluated over embedded devices and using the CANoe software tool for simulation, and it is noticed that it relatively takes more computation time. However, it is required to reduce the computation time during the authentication process, as CAN is used in CAVs and other safety-critical applications.

A significant problem of session key agreement over AUTomotive Open System ARchitecture (AUTOSAR) compliance is not resolved effectively, even though various message authentication protocols are proposed for CAN communications. An AUTOSAR-compliant key management architecture is proposed in \cite{Xiao2020ACSEC} by considering practical requirements for the automotive system. Further, a baseline Session Key Distribution protoCol (SKDC) is designed to provide various security functionalities, and a new Secret Sharing key Transfer (SSKT) protocol is proposed to achieve better communication efficiency results. The implementation of Arduino IDE and the CAN Bus Shield library confirms that SSKT provides better computation and communication results.

CAN FD is advantageous for data transmission in IVN because of its bit-rate capacity (of 8 Mbps) and payload size (of 64 bytes). However, it is vulnerable to masquerade attacks due to the unavailability of adequate authentication protocols. In \cite{Xie2020ITSCANFD}, a two-stage scheme is proposed with two algorithms for security improvements for CAN FD communications. The first stage is performed to get the lower bound of an in-vehicle application by omitting most sequences through a quick sequence abandoning algorithm. Moreover, the laxity interval values are obtained from the lower bound to the deadline. In the second stage, the round accumulation algorithm is executed to improve the security by using MACs to CAN FD messages. The performance analysis results show that the proposed scheme is suitable for enhancing IVN communications security.

In CAVs, it is necessary to protect the TIS, ECUs, and OBD-II ports against message spoofing attacks due to their importance in IVN. A CAN bus authentication scheme is proposed in \cite{Xiao2021TIFS} that uses message physical layer features, i.e., message arrival intervals and signal voltages, applying a reinforcement learning approach to select the authentication mode and parameter. The proposed scheme achieves better authentication accuracy without modifying the CAN bus protocol's ECU parts. Moreover, a deep learning-based authentication scheme is proposed using a hierarchical structure and two deep neural networks, reducing the exploration time and compressing the high-dimensional state space with fully exploiting physical layer features. Thus, it provides superior authentication efficiency over the CAN bus, as it is verified through a test-bed setup with embedded devices.

Recent security experiments demonstrated the possibility of illegal access to car functionalities and vehicle theft, making modern vehicles vulnerable in different ways. To deal with these challenges, secure access and feature activation scheme is proposed in \cite{Plappert2021CCS} based on TPM 2.0 (acting as a trust anchor in a vehicle), and thereby, it provides a fine-granular authorization mechanism. Moreover, this proposed system can protect against potential security attacks in automotive scenarios. The experimental results on Raspberry Pi show that it can achieve good performance results, but it could be improved for better performance efficiency to enable superior performance in automotive systems.

The secure exchange of cryptographic keys between ECUs is a significant challenge for secure IVN communications. In \cite{Musuroi2021TVT}, authors evaluated the key exchange protocol based on a standardized National Institute of Standards and Technology (NIST) elliptic curve and Four$Q$ curve of the Diffie-Hellman. The implementation results of these protocols over Infineon and ARM core processor platforms show effective performance for CAN and CAN FD. It is also noticed that the computation time is more crucial than bandwidth, as the execution time of the elliptic curve is relatively high.

Attackers can launch masquerade, suspension, and injection attacks on the CAN bus architecture due to the lack of appropriate built-in authentication and encryption mechanisms, resulting in life-damaging consequences. A Transmitter Authentication scheme in CAN (TACAN) is proposed in \cite{Ying2021TDSC} to offer secure authentication between deployed ECUs over the CAN bus architecture through three different covert channels (inter-arrival time-based, least significant bit-based, and hybrid). Further, TACAN can be implemented without CAN protocol modifications and communication overheads. The extensive experimental results on Chevrolet Camaro 2016 and Toyota Camry 2010 datasets demonstrate that TACAN effectively detects CAN bus attacks and achieves better results when evaluating bit error and throughput performance parameters.

Table \ref{MessageAuthenticationComparision} presents a comparative study of state-of-the-art message authentication-based ADPS. This table gives a better overview of relevant ADPS solutions to understand their efficacy in various features.

\begin{table*}[!h]
\centering
\caption{Comparison of State-of-the-art Message Authentication-based Protocols among Different Attributes}
\label{MessageAuthenticationComparision}
\scalebox{0.70}{		
\begin{tabular}{|p{1.2in}|c|p{2.3in}|P{1.1in}|P{0.85in}|P{0.8in}|p{2.1in}|}  \hline
\textbf{Scheme} & \textbf{Year} & \textbf{Objective(s)} & \textbf{Protection for Attacks} & \textbf{Impact on System/Device} & \textbf{Computational Resource Requirement} & \textbf{Limitations/Scope of Enhancement} \\ \hline

Woo et al. \cite{Woo2014ITS} & 2014 & Provide protection against long-range wireless attacks; & Replay, Injection; & ECUs; & High; & Weak against authentication attacks; \\ \hline

Radu and Garcia \cite{Radu2016ESORICS} & 2016 & To provide mutual authentication between ECUs & Modification, Impersonation, Chosen message, Injection; & ECUs, CAN bus; & --- & Require to analyze performance results in addition to security analysis; \\ \hline
Poudel and Munir \cite{Poudel2018TDSC} & 2018 & Design an ECU architecture by integrating security and dependability attributes with low computational resources overhead; & Injection, Eavesdropping; & ECUs; & Low & Secure storage/generation/distribution of keys, authentication of ECUs, and privacy regulations; \\ \hline

Groza et al. \cite{Groza2019Access} & 2019 &Provide secure key exchanges between two CAN components; & Impersonation; & ECUs, CAN; & Average & Not enough security strength to resist current cyberattacks; \\ \hline

Joo et al. \cite{Joo2020NDSS} & 2020 & Attacks detection on keyless entry systems, exploiting the RF-fingerprint technique; & Relay, Injection; & Keyless entry systems; & Average & Identification of a variety of security threats in keyless entry systems;\\ \hline
Palaniswamy et al. \cite{Palaniswamy2020TIFS} & 2020 & Provide secure key computations and update for IVN operations; & Impersonation, replay, and man-in-the-middle; & ECUs, CAN; & Average & Session key availability to compromised ECUs;\\ \hline
Jo et al. \cite{Jo2020TVT} & 2020 & Provide authentication protocol to resist masquerade attacks without utilizing fully network capacity and requiring hardware changes & Masquerade, Bus-off, Replay, Fabrication; & CAN; & Average & Require to improve performance in different parameters, i.e., the waiting time; \\ \hline
Xiao et al. \cite{Xiao2020ACSEC} & 2020 & Effective session key establishment and secure distribution; & Impersonation, Replay; & CAN/CAN-FD bus; & Average & Possibility to minimize the resource requirement;\\ \hline
Xie et al. \cite{Xie2020ITSCANFD} & 2020 & Enhancement of security in CAN-FD; & Masquerade; & CAN; & Low & New designs for non-parallel IVN applications with security enhancements; \\ \hline
Xiao et al. \cite{Xiao2021TIFS} & 2021 & Enhance authentication accuracy without changing the CAN bus protocol or the ECUs and requiring knowledge of the spoofing model; & Impersonation, Replay, Bus-off, Man-in-the-middle; & ECUs, CAN bus; & --- & Need to improve performance results and provide protection for the monitor and ECUs; \\ \hline

Plappert et al. \cite{Plappert2021CCS} & 2021 & Provide a secure access and feature activation system to protect potential security threats; & Injection, Modification, Replay, Eavesdropping, Man-in-the-middle; & Trusted Trusted Module (TPM); & Average & Possibility to minimize the overhead and improve TPM 2.0-inherent policy;  \\ \hline
Musuroi et al. \cite{Musuroi2021TVT} & 2021 & To securely exchange cryptographic keys between ECUs with fast computations; & Impersonation, Replay; & CAN bus; & Average; & Require to give attention on the group key exchange method for high performance results;\\ \hline
Ying et al. \cite{Ying2021TDSC} & 2021 & Provides secure authentication for connected ECUs over CAN; & Injection, Impersonation; & ECUs, CAN; & Average; & Possibility to improve performance results for quick and better protection against attacks; \\ \hline

\end{tabular}}
\end{table*}

\subsubsection{Other Approaches}
Some other approaches are helpful in detecting various security attacks and providing protection against them.

\noindent \textbf{\underline{Anti Analysis}:} 
Attestation is the mechanism in which software verifies the authenticity and integrity of the hardware and software of a device. In today's CAVs, ECUs use flash memory that allows authorized entities to update or flash a new version of the firmware. Although firmware updates, especially common in CAVs, fix known bugs and security holes in the software, it increases the attack surface. Therefore, knowing when the system's integrity has been compromised is crucial, which can be achieved by using cryptographically secure techniques such as firmware attestation, MAC, and hash-value authentication. The firmware attestation scheme is a challenge and response type of protocol. Two main entities are involved in the attestation process, a challenger (the attester) and a respondent (the ECU being attested) \cite{Rawat2021Decentralized}.

The most important feature of anti-analysis-based ADPS is the integrity of the firmware, which allows each ECU to learn about the security stance of other ECUs in the vehicle. Furthermore, a decentralized attestation process is more robust and can independently attest to the state of the whole vehicle. However, anti-analysis-based IDS is ineffective against attacks on the program without affecting the state of the firmware. It includes attacks in the current memory program and on the trusted hardware, affecting the attestation process's trustability. 

\noindent \textbf{\underline{Post Protection}:} 
Firmware Over-The-Air (FOTA) update is the process of distributing new firmware via the wireless medium (i.e., Wi-Fi and cellular network) to update the application that runs on top of the operating system. The updates usually come with software fixes, new features, and enhancements for the vehicles. This process updates the whole software stack and replaces the operating system and application. FOTA is especially critical for CAVs, as they are constantly connected to the external networks and need to be updated fast to deal with new threats and environments regularly. A secure firmware over-the-air update can prevent the firmware from compromising \cite{Jo2021ITS}.

\noindent \textbf{\underline{Fuzzing}:} 
Fuzzing is a security testing technique that attempts to find software bugs by injecting randomly generated valid and invalid inputs into a program. A fuzzer software is usually used to automatically create a set of test values. A normal program would expect to receive structured inputs, and fuzzing stress tests the application to create unexpected behavior or crashes. CAN can expose unknown vulnerabilities in the ECU software while fuzzing is applied on CAN traffic \cite{Huang2018ATGARES}.

Fuzzing on ECUs is more challenging due to car manufacturers' different proprietary CAN databases. The CAN database is specified in the Database Container (DBC) format file, a text file containing information for decoding raw CAN bus data to "physical values." While black-box methods such as brute-force and random search can work without the CAN database, they are inefficient due to the infinite number of possible inputs. Fuzzing detects loopholes in software reliably without false positives, increasing the robustness of car software. With the advent of CAVs, fuzzing will be more important as more software is deployed and the connected vehicles suffer similar security vulnerabilities to other computer-based network systems. In addition, fuzzing can help discover vehicle system functions that car manufacturers may not know \cite{Choi2021TVTDBC}.

\begin{figure*}[!h]
\centering
{\includegraphics[width=0.95\textwidth]{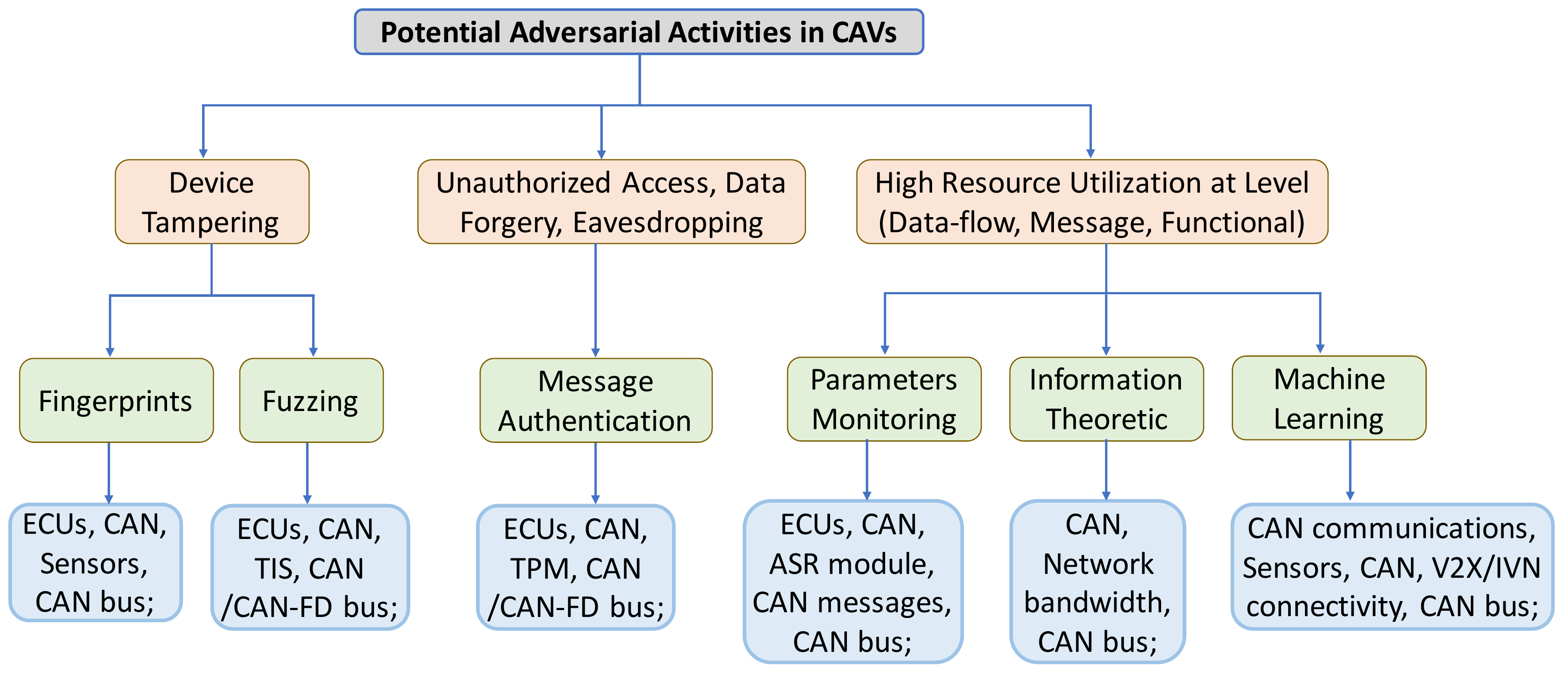}}
\caption{Correlation of Different ADPS Methods with Various Affecting IVN Components}
\label{AdversarialActivities}
\end{figure*}

Figure \ref{AdversarialActivities} displays potential adversarial activities (at the first level) that can be carried out to damage CAVs. Various attack detection and prevention categories are linked to specific malicious exploits that help to understand their effectiveness in identifying intrusions and prevent against them. Different autonomous vehicle components are presented under each ADPS category to protect them from system damage. This graphical presentation gives a better understanding of the association among adversary interests, ADPS, and relevant vehicle components. This, in turn, helps researchers in developing security solutions. Researchers have developed various security solutions using different ADPS categories to improve the automotive system's strengths against vulnerable activities. However, technological development imposes advanced threats on the automotive systems, leading to new security and performance challenges.

\section{Research Directions for CAVs}
Modern vehicles are connected with external interfaces, several software modules, and many ECUs via OBD-II. That exposes CAVs to malicious activities with conventional and new security threats. The market for CAVs is rapidly increasing to provide more advanced transportation services and comfortable journeys. Hence, it has become essential to detect security vulnerabilities and faults in CAVs; otherwise, it can create chaos on the road, causing undesired consequences, human life risk, or infrastructure damage. Besides, there are other approaches (i.e., keyless entry system, telematics, DSRC/Bluetooth/LTE/Wi-Fi communication technologies, and Global Positioning System) through which adversaries can target the automotive system for susceptible activities in CAVs. The keyless entry system has received the highest attention from adversaries for malicious actions by performing signal relay attacks. We, therefore, discuss key research problems and open challenges for ADPS of CAVs. 

\subsection{Systematic Fuzz Testing Methodologies} 
Datasets with normal and attack scenarios are commonly used to identify security threats and validate novel attack detection techniques. However, limited research works are available on the collection and validation of the attacks data \cite{Aliwa2021CSUR}. Such realistic datasets are valuable assets to the research community to continuously improve the resilience of security assessment solutions for CAVs and accurately measure the performance of attack detection strategies. Since the market of CAVs is increasing rapidly to enable society with advanced transportation services and applications, there is an immediate need to develop systematic fuzzing-based security testing techniques. Such fuzz testing methodologies may facilitate continuous testing for a variety of attacks to realize the resilience of CAV systems and evaluate the effectiveness of attack detection and prevention approaches in a real-time environment based on different performance measurement parameters, e.g., accuracy, timing, sensitivity, etc. Moreover, the progress in fuzzing methodologies has opened new avenues to discover unforeseen (zero-day) attacks on CAVs. Such is crucial to fine-tune the automotive security systems before deployment. 

Fuzzing approaches can be classified into Blackbox, Greybox, and Whitebox. It is not feasible to use any form of Greybox fuzzing approaches \cite{Pham2019TSE}, as such approaches require instrumenting the ECU code. Whitebox approaches, e.g., symbolic execution \cite{KLEEFuzzing} is also not applicable for fuzzing commercial ECUs, as commercial ECUs are closed source. Existing Blackbox fuzzing approaches \cite{RadamsaFuzzing}, \cite{Nishimura2016ICVES} are unlikely to be effective, as such techniques (i) do not learn from previous fuzzing campaigns, or (ii) are limited in terms of structured input generation, and these are important aspects for effective protocol fuzzing. Systematic Blackbox fuzzing, which will generate structured inputs according to the targeted protocol and learn from the fuzzing campaigns to automatically evolve the fuzzing process, is likely to be effective and practical for fuzzing components of CAVs. This can be accomplished by maximizing the explored protocol features to uncover new vulnerabilities.

\subsection{Device-based Novel Attack Detection Mechanisms}
CAV relies on a large number of multiple sensors, including cameras, radars, and LIDARs, enabling more accurate data results for worry-free journeys. However, these sensors enable adversaries for additional attack surfaces to launch sensor-based attacks (such as spoofing, eavesdropping, and jamming) on the vehicle's self-driving automated control system~\cite{Pham2021CS}. Such additional attack surfaces may lead to information leakage, false sensory data injection, DoS, and transmission of malicious commands in the IVN~\cite{Sikder2021CSTSensorSurvey}. Since CAVs are highly mobile nodes and gather data from various sensors to perform different operations with limited resources, detecting malicious or faulty sensor nodes is challenging. Advanced attack detection systems that combine groundbreaking techniques (such as sensor fusion and machine learning with the abundance of information generated by CAVs) should be developed to detect sensor-based attacks effectively.

\subsection{Compromised ECU Identification}
Current security solutions can provide a specific level of security robustness over the CAN bus architecture to protect from forgery attacks (that are launched to disrupt the communication channel or automotive data). Available solutions are limited in scope (for data protection and communication channel) and can withstand specific security attacks only. Thus, it is difficult to identify susceptible activities when a compromised part (i.e., ECU) launches attacks on the CAN bus architecture. To find the source of attacks, protocols based on electrical signal characteristics of ECUs are proposed~\cite{EASI2020NDSS}, \cite{Cho2017CCS}, \cite{Choi2018TIFS}, \cite{Pese2019CCS}, \cite{Wang2020RAID}. Such solutions may not realize whether the source is already compromised or not due to the change in IVN environmental circumstances. ECUs can be compromised in two ways: (i) exposed ECU is mounted, and (ii) ECU is compromised after the installation. Zero-trust-based multi-factor authentication protocols should be implemented by involving multiple entities during the deployment of ECUs to avert the first possibility of ECU compromising. For effective attack identification from compromised ECUs, lightweight security protocols should be developed to protect the system from compromised ECUs quickly.

\subsection{Lightweight Security Protocols for IVN}
CAN is not enabled with an in-built authentication and encryption mechanism to protect from forgery attacks over the CAN bus architecture. Therefore, researchers have focused on addressing the issue of forged communications by developing cryptographic-based security solutions. Hardware-based cryptography methods can improve the security level to meet the real-time needs of CAVs. However, the high implementation cost, the compatibility with the existing infrastructure, and system modifications are important challenges in satisfying security requirements. Software-based cryptography methods can be applied and do not require changes in the CAN bus architecture. However, the computation and communication overhead on the payload increase the requirement of additional computing capabilities on resource-constrained automotive systems, leading to a time-consuming process~\cite{Aliwa2021CSUR}, \cite{EASI2020NDSS}, \cite{Woo2014ITS}, \cite{Palaniswamy2020TIFS}. Researchers have developed various security schemes to provide security using different cryptographic primitives, but most of them require more computation cost and communication overhead. To reduce the network delay in ITS applications and services, it is required to upgrade the vehicular network with the latest communication technology. 6G-enabled vehicular networks can offer better network connectivity to minimize the communication latency, but the system consumes more energy in performing computing and communication operations \cite{Zhu2022ITII6G}, \cite{Xu2022IoT6G}. Therefore, the key challenge is to design lightweight security protocols for CAN-based communications with low latency. This is to perform necessary operations quickly with limited computing power and provide an adequate level of security to protect the automated system from various security attacks.

\subsection{Malware Code Resilient CAVs}
CAVs are configured with IoT and embedded devices to execute in-vehicle and outside network operations to make better decisions. These devices are very limited in security features to avert various threats~\cite{Babun2021CNIoTSurvey}, and thereby they are the major targets of adversaries to launch traditional and new security attacks through malware codes \cite{ElRewini2019VC}. Since ECUs are connected to external sources through a gateway, real-time malware scanning can be applied at the gateway. However, the need for excessive computing power is raised for a gateway, which might not detect all malware codes with its limited on-board resources. Furthermore, it is tough to identify malware amongst the high number of associated ECUs in CAVs \cite{Zhang2014IoTMalware}, \cite{Elkhail2021Access}. Thus, it opens an opportunity for adversaries to send malicious payloads (through SQL injection vulnerability) to perform susceptible activities over the IVN, leading the automotive system to unanticipated situations and severe consequences. Hence, it is adequate to design the automotive system with malware code detection and protection to reduce the impact of security exposures and vulnerabilities.

\subsection{Control-Oriented Techniques for 6G-enabled Infrastructure}
CAVs communicate and transfer high-cost computations to the infrastructure through V2X communication technology for rich data inputs and minimizing the requirement of computational resources at the CAV level, thus improving the effectiveness of CAVs with more accurate operations. Integrating 6G communication technology with the vehicular network for high throughput, better decision-making abilities, and reduced latency is necessary to exchange various computations productively and their outcomes between the infrastructure and vehicles \cite{Xu2022IoT6G}. Vehicular infrastructure is susceptible to cyber threats, including malware, weak access control, and limited security features over the firmware process \cite{Acharya2020IEEEAccess}. Blockchain, molecular/Terahertz/quantum/Visible light communications, and AI technologies are important in 6G communications. However, they are vulnerable to malicious behavior, access control/authentication/integrity attacks, eavesdropping, and data transmission exposure, creating security and privacy issues \cite{Wang2020DCN}. Thus, an adversary can launch replay, bogus message, modification, blackhole, and wormhole attacks in the IVN. Thus, automated driving system operations are significantly impacted, disrupting the overall performance of a platoon of CAVs. To improve the safety and security of autonomous driving systems, advanced and robust vehicular control frameworks should be developed to withstand traditional and new cyberattacks~\cite{Aliwa2021CSUR}. Current research has mainly focused on the prevention and defense techniques for CAVs, but it is also required to emphasize control and recovery strategies that can support damaged infrastructure (i.e., RSUs and cloud servers) to recover from unexpected incidents effectively and security vulnerabilities \cite{Mughal2020VC}. CAVs automatically execute various vehicle operations (considering the available information (from in-vehicle components) or obtained data from the infrastructure) without (or minor) human intervention while on the move. It is thereby necessary to quickly restore the system from damaged conditions and perform different operations by following legal system procedures. Control-oriented techniques manage the automated control and recovery from attacks that can reduce the damage level in CAVs. Hence, the research area of control-oriented techniques should be explored to create effective resilient and recovery strategies that can mitigate such network attacks in the IVN.

\subsection{Recognition of Adversarial AI Attacks}
Researchers suggested various machine learning-based models to detect security attacks in CAVs, and these models mainly work based on the collected data through installed devices. However, there are demonstrations that if pixel values of an input image are altered, then the model can produce erroneous results, and the understanding of images is successful under certain conditions only \cite{Nguyen2015CVPR}, \cite{Qayyum2020IEEECOMST}. Besides, various mechanisms are trained to understand "patches", and they can be imposed on an object to mislead detectors and classifiers \cite{Chen2019CybersecurityAI}, \cite{Kong2020CVPR}, \cite{Nassi2020CCS}, \cite{Wang2021CCS}. In such cases, the trained model cannot detect objects even though they are available on the way to CAVs, or they can come closer to CAVs \cite{Deng2020Percom}. Thus, an attacker may cause significant damage by launching adversarial attacks on reinforcement learning mechanisms. Therefore, it is essential to develop reliable machine learning-based attack detection systems for CAVs.

\subsection{Trustworthy Fog-enabled Vehicular Networks}
Autonomous vehicles produce around 20 GB of data per hour, and they are required to collect, analyze, process, and aggregate the gathered data before using relevant information in vehicular applications and services that may delay emergency and navigation services \cite{Hussain2018CSAT}. The concept of vehicular fog computing was introduced to reduce the computational overhead at resource-constrained devices (i.e., vehicles) by executing high-cost operations at fog devices (i.e., edge servers and/or RSUs). In this, vehicles collect relevant data from sensors (installed in a vehicle) and transfer gathered data to the fog devices for data analysis (to provide better ITS services to vehicle travelers). Fog devices can then deliver such information to CAVs, which can be used to make timely, on-road decisions \cite{Pereira2019FGCS}, \cite{Sun2019JNCA}. Though fog-enabled vehicular frameworks minimize the computational resource requirements at CAVs, there are still significant security and privacy challenges: trustworthiness of delivered data (from vehicles to fog devices), secure data transfer, off-loading of tasks, and on-time information availability (to CAVs) in highly mobile environments \cite{Ni2017ICM}, \cite{Kang2017ITS}, \cite{Nkenyereye2021}. Connected vehicle technology is focused to enable CAVs safer, faster, and more efficient, but the availability of erroneous information to CAVs may lead to accidental consequences \cite{Soleymani2021VC}, \cite{Liu2021MNAA}, \cite{Singh2021JPDC}. Therefore, developing reliable and efficient attack detection and prevention mechanisms for fog-based vehicular networks is vital.

\subsection{Dependable Digital Twin-based Automated Driving Systems}
Digital Twin (DT) is one of the most cutting-edge technologies of Industry 4.0 that is developed as a virtual representation of physical entities with simulation proficiency to predict and optimize states, functionality, and configurations. DT synchronizes the mapping with the physical object that is useful in real-time monitoring, object management, data analytics, maintenance strategies, and risk estimation \cite{Wu2021IoT}. Therefore, the integration of DT with CAVs is worthwhile to identify potential issues in automotive driving systems, get real-time feedback on automated operations, control vehicles in uncertain situations, and realize performance improvements \cite{Jones2020CIRP}, \cite{Li2021TVT}, which is beneficial in the development of accurate attacks detection and prevention systems for CAVs. As DT modeling depends on the data (received from the synchronized physical object over V2X connectivity), it is important to securely and efficiently deliver data between digital and physical spaces \cite{Guo2022DCN}. However, providing security over V2X technology is challenging due to the shared communication link  \cite{ElRewini2019VC}. Besides, the communication delay is a vital factor for DT-enabled CAVs as the unavailability of necessary data on-time can put CAVs in unexpected situations, creating user and road safety issues \cite{Alcaraz2022CSAT}. It is essential to develop dependable DT-based mechanisms for CAVs for effective services on the road.

\begin{figure*}[!h]
\centering
{\includegraphics[width=1.0\textwidth]{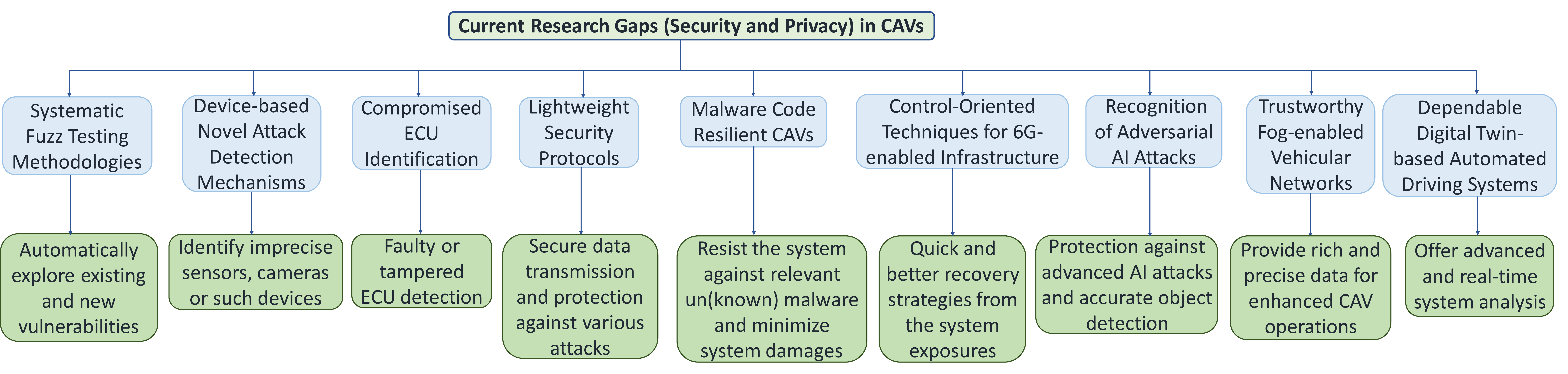}}
\caption{The Outline of Important Research Directions for CAVs with Potential Features}
\label{ResearchDirections}
\end{figure*}

Figure \ref{ResearchDirections} shows a graphical presentation of important security and privacy research directions in connected and automated driving systems. As these research gaps are crucial in the development of reliable CAVs for the benefit of society, it is required to focus on these challenges to enable vehicle travelers with safe, secure, and intelligent vehicles in the near future. Figure \ref{ResearchDirections} also displays that developing adequate solutions for these research problems can offer various features to improve the states, operations, and functionality of CAVs.

\section{Conclusions}
This survey article gives an overview of CAVs in different aspects. Considering the significance, important applications, and mobility nature of CAVs, we have discussed vital security and privacy properties as well as performance evaluation parameters to understand their importance in CAVs. Moreover, a variety of attacks are briefly explained, and their possible countermeasures are discussed. Such potential attacks significantly impact the automotive system of CAVs and can produce unexpected consequences. We have extensively reviewed different categories of ADPS and have systematically studied recent IVN solutions to classify them under the category of attack detection and protection. To quickly provide in-depth knowledge about the current research status on ADPS approaches, we present a comparative summary of relevant methods under each category by providing their key contributions, features, and scope for enhancement. We hope this survey will provide a strong base to study recent ADPS solutions and research directions for new and more appropriate techniques to achieve better security and performance efficiency.

\section*{Acknowledgement}
This work is supported by the grant from Land Transport Authority (LTA), Singapore (LTA-UMGC-L011). We thank LTA colleagues and project team members for their helpful input.


\begin{thebibliography}{100}

\bibitem{Kim2021CS}  
Kim, K., Kim, J. S., Jeong, S., Park, J. H., and Kim, H. K. (2021). Cybersecurity for autonomous vehicles: Review of attacks and defense. Computers \& Security, 102150, pp. 1-27.

\bibitem{Sun2021ITS}
Sun, X., Yu, F. R., and Zhang, P. (2021). A Survey on Cyber-Security of Connected and Autonomous Vehicles (CAVs). IEEE Transactions on Intelligent Transportation Systems, pp. 1-20.

\bibitem{SAEAutomationLevels2021}
Taxonomy and Definitions for Terms Related to Driving Automation Systems for On-Road Motor Vehicles. [Online] \url{https://www.sae.org/standards/content/j3016_202104/}

\bibitem{Wang2018CST}
Wang, J., Liu, J., and Kato, N. (2018). Networking and communications in autonomous driving: A survey. IEEE Communications Surveys \& Tutorials, 21(2), 1243-1274.

\bibitem{Zeng2016ICST}
Zeng, W., Khalid, M. A., and Chowdhury, S. (2016). In-vehicle networks outlook: Achievements and challenges. IEEE Communications Surveys \& Tutorials, 18(3), 1552-1571.

\bibitem{Huang2017ISPEC}
Huang, T., Zhou, J., Wang, Y., and Cheng, A. (2017, December). On the security of in-vehicle hybrid network: Status and challenges. In International Conference on Information Security Practice and Experience (pp. 621-637). Springer, Cham.

\bibitem{Aliwa2021CSUR}
Aliwa, E., Rana, O., Perera, C., and Burnap, P. (2021). Cyberattacks and countermeasures for in-vehicle networks. ACM Computing Surveys (CSUR), 54(1), 1-37.
 
\bibitem{EASI2020NDSS}
Kneib, M., Schell, O., and Huth, C. (2020, February). EASI: Edge-Based Sender Identification on Resource-Constrained Platforms for Automotive Networks. In 27th Network and Distributed System Security Symposium (NDSS), pp. 1-16.


\bibitem{CANXL2020}
CAN XL is knocking on the door. [Online] \url{https://www.can-cia.org/news/cia-in-action/view/can-xl-is-knocking-on-the-door/}

\bibitem{Jo2021ITS}
Jo, H. J., and Choi, W. (2021). A Survey of Attacks on Controller Area Networks and Corresponding Countermeasures. IEEE Transactions on Intelligent Transportation Systems, pp. 1-19.

\bibitem{FutureMobility2015Deloitte}
The future of mobility. [Online]
\url{https://www2.deloitte.com/us/en/insights/focus/future-of-mobility/transportation-technology.html}

\bibitem{FutureTransport2021NSW}
Future Transport Technology Roadmap 2021-2024. [Online]
\url{https://future.transport.nsw.gov.au/technology/technology-roadmap}

\bibitem{CAVEnergyConsumption2017EIA}
Study of the Potential Energy Consumption Impacts of Connected and Automated Vehicles. [Online] \url{https://www.eia.gov/analysis/studies/transportation/automated/pdf/automated_vehicles.pdf}

\bibitem{ECommerce2021Jilt}
The eCommerce decade: How the 2010s changed online shopping. [Online]
\url{https://jilt.com/blog/decade-ecommerce-2010s/}

\bibitem{ParcelDelivery2016McKinsey}
How customer demands are reshaping last-mile delivery. [Online] \url{https://www.mckinsey.com/industries/travel-logistics-and-infrastructure/our-insights/how-customer-demands-are-reshaping-last-mile-delivery}

\bibitem{Woo2014ITS}
Woo, S., Jo, H. J., and Lee, D. H. (2014). A practical wireless attack on the connected car and security protocol for in-vehicle CAN. IEEE Transactions on intelligent transportation systems, 16(2), 993-1006.

\bibitem{Petit2014ITS}
Petit, J., and Shladover, S. E. (2014). Potential cyberattacks on automated vehicles. IEEE Transactions on Intelligent transportation systems, 16(2), 546-556.

\bibitem{Humayed2017IoT}
Humayed, A., Lin, J., Li, F., and Luo, B. (2017). Cyber-physical systems security—A survey. IEEE Internet of Things Journal, 4(6), 1802-1831.

\bibitem{Checkoway2011USENIX}
Checkoway, S., McCoy, D., Kantor, B., Anderson, D., Shacham, H., Savage, S., Koscher, K., Czeskis, A., Roesner, F., Kohno, T. (2011, August). Comprehensive experimental analyses of automotive attack surfaces. In USENIX Security Symposium, Vol. 4, No. 447-462.

\bibitem{Miller2015BlackHat}
Miller, C., and Valasek, C. (2015). Remote exploitation of an unaltered passenger vehicle. Black Hat USA, pp. 1-91.

\bibitem{Cai2019BlackHat}
Cai, Z., Wang, A., Zhang, W., Gruffke, M., and Schweppe, H. (2019). 0-days \& mitigations: roadways to exploit and secure connected BMW cars. Black Hat USA, 1-37.
 
\bibitem{Bloom2017SOUPS}
Bloom, C., Tan, J., Ramjohn, J., and Bauer, L. (2017). Self-driving cars and data collection: Privacy perceptions of networked autonomous vehicles. In 13th Symposium on Usable Privacy and Security (SOUPS), pp. 357-375. 
 
\bibitem{Joy2017ICCCN}
Joy, J., and Gerla, M. (2017, July). Internet of vehicles and autonomous connected car-privacy and security issues. In 26th International Conference on Computer Communication and Networks (ICCCN), pp. 1-9.

\bibitem{Fu2020IWC}
Fu, Y., Yu, F. R., Li, C., Luan, T. H., and Zhang, Y. (2020). Vehicular blockchain-based collective learning for connected and autonomous vehicles. IEEE Wireless Communications, 27(2), 197-203.

\bibitem{Lim2018Energies}
Lim, H. S. M., and Taeihagh, A. (2018). Autonomous vehicles for smart and sustainable cities: An in-depth exploration of privacy and cybersecurity implications. Energies, 11(5), 1062.

\bibitem{Wu2019ITS}
Wu, W., Li, R., Xie, G., An, J., Bai, Y., Zhou, J., and Li, K. (2019). A survey of intrusion detection for in-vehicle networks. IEEE Transactions on Intelligent Transportation Systems, 21(3), 919-933.

\bibitem{Pham2021CS}
Pham, M., and Xiong, K. (2021). A survey on security attacks and defense techniques for connected and autonomous vehicles. Computers \& Security, 102269, pp. 1-29.

\bibitem{GarciaTeodoro2009CS}
Garcia-Teodoro, P., Diaz-Verdejo, J., Macia-Fernandez, G., and Vazquez, E. (2009). Anomaly-based network intrusion detection: Techniques, systems and challenges. computers \& security, 28(1-2), 18-28.


\bibitem{Lokman2019EJWCN}
Lokman, S. F., Othman, A. T., and Abu-Bakar, M. H. (2019). Intrusion detection system for automotive Controller Area Network (CAN) bus system: a review. EURASIP Journal on Wireless Communications and Networking, 2019(1), 1-17.

\bibitem{Frassinelli2020IEEESP}
Frassinelli, D., Park, S., and Nurnberger, S. (2020, May). I know where you parked last summer: Automated reverse engineering and privacy analysis of modern cars. In 2020 IEEE Symposium on Security and Privacy (SP) (pp. 1401-1415). IEEE.

\bibitem{Li2019WCSideChannel}
Li, Y., Luo, Q., Liu, J., Guo, H., and Kato, N. (2019). TSP security in intelligent and connected vehicles: Challenges and solutions. IEEE Wireless Communications, 26(3), 125-131.

\bibitem{Cho2016USENIX}
Cho, K. T., and Shin, K. G. (2016). Fingerprinting electronic control units for vehicle intrusion detection. In 25th {USENIX} Security Symposium ({USENIX} Security 16) (pp. 911-927).

\bibitem{Cho2017CCS}
Cho, K. T., and Shin, K. G. (2017, October). Viden: Attacker identification on in-vehicle networks. In Proceedings of the 2017 ACM SIGSAC Conference on Computer and Communications Security (pp. 1109-1123).


\bibitem{Choi2018TIFS}
Choi, W., Joo, K., Jo, H. J., Park, M. C., and Lee, D. H. (2018). Voltageids: Low-level communication characteristics for automotive intrusion detection system. IEEE Transactions on Information Forensics and Security, 13(8), 2114-2129.

\bibitem{Kneib2018CCS}
Kneib, M., and Huth, C. (2018, October). Scission: Signal characteristic-based sender identification and intrusion detection in automotive networks. In Proceedings of the 2018 ACM SIGSAC Conference on Computer and Communications Security (pp. 787-800).

\bibitem{Foruhandeh2019ACSAC}
Foruhandeh, M., Man, Y., Gerdes, R., Li, M., and Chantem, T. (2019, December). SIMPLE: Single-frame based physical layer identification for intrusion detection and prevention on in-vehicle networks. In Proceedings of the 35th Annual Computer Security Applications Conference (pp. 229-244).

\bibitem{Sun2021TIFS}
Sun, Z., Balakrishnan, S., Su, L., Bhuyan, A., Wang, P., and Qiao, C. (2021). Who Is in Control? Practical Physical Layer Attack and Defense for mmWave-Based Sensing in Autonomous Vehicles. IEEE Transactions on Information Forensics and Security, 16, 3199-3214.

\bibitem{Bhatia2021NDSS}
Bhatia, R., Kumar, V., Serag, K., Celik, Z. B., Payer, M., and Xu, D. (2021, February). Evading voltage-based intrusion detection on automotive CAN. In Network and Distributed System Security Symposium (NDSS).

\bibitem{Cho2016CCS}
Cho, K. T., and Shin, K. G. (2016, October). Error handling of in-vehicle networks makes them vulnerable. In Proceedings of the 2016 ACM SIGSAC Conference on Computer and Communications Security (pp. 1044-1055).

\bibitem{Kulandaivel2019USENIX}
Kulandaivel, S., Goyal, T., Agrawal, A. K., and Sekar, V. (2019). Canvas: Fast and inexpensive automotive network mapping. In 28th {USENIX} Security Symposium ({USENIX} Security 19) (pp. 389-405).

\bibitem{Sun2019Access}
Sun, H., Lee, S. Y., Joo, K., Jin, H., and Lee, D. H. (2019). Catch ID if you CAN: Dynamic ID virtualization mechanism for the controller area network. IEEE Access, 7, 158237-158249.

\bibitem{Pese2019CCS}
Pesé, M. D., Stacer, T., Campos, C. A., Newberry, E., Chen, D., and Shin, K. G. (2019, November). LibreCAN: Automated CAN message translator. In Proceedings of the 2019 ACM SIGSAC Conference on Computer and Communications Security (pp. 2283-2300).

\bibitem{Ying2019TIFS}
Ying, X., Sagong, S. U., Clark, A., Bushnell, L., and Poovendran, R. (2019). Shape of the cloak: Formal analysis of clock skew-based intrusion detection system in controller area networks. IEEE Transactions on Information Forensics and Security, 14(9), 2300-2314.

\bibitem{Othmane2020TDSC}
Othmane, L. B., Dhulipala, L., Abdelkhalek, M., Multari, N., and Govindarasu, M. (2020). On the performance of detecting injection of fabricated messages into the can bus. IEEE Transactions on Dependable and Secure Computing.

\bibitem{Olufowobi2020TVT}
Olufowobi, H., Young, C., Zambreno, J., and Bloom, G. (2019). Saiducant: Specification-based automotive intrusion detection using controller area network (can) timing. IEEE Transactions on Vehicular Technology, 69(2), 1484-1494.

\bibitem{Wang2020RAID}
Wang, S., Cao, J., Sun, K., and Li, Q. (2020). {SIEVE}: Secure In-Vehicle Automatic Speech Recognition Systems. In 23rd International Symposium on Research in Attacks, Intrusions and Defenses ({RAID} 2020) (pp. 365-379).

\bibitem{Murvay2020Access}
Murvay, P. S., and Groza, B. (2020). TIDAL-CAN: Differential timing based intrusion detection and localization for controller area network. IEEE Access, 8, 68895-68912.

\bibitem{Olufowobi2019Anomaly}
Olufowobi, H., Ezeobi, U., Muhati, E., Robinson, G., Young, C., Zambreno, J., and Bloom, G. (2019). Anomaly detection approach using adaptive cumulative sum algorithm for controller area network. Proceedings of the ACM Workshop on Automotive Cybersecurity, 25–30.

\bibitem{Bozdal2021}
Bozdal, M., Samie, M., and Jennions, I. K. (2021). WINDS: A Wavelet-Based Intrusion Detection System for Controller Area Network (CAN). IEEE Access, 9, 58621-58633.

\bibitem{Groza2021TIFS}
Groza, B., Popa, L., and Murvay, P. S. (2021). CANTO-Covert AutheNtication with Timing channels over Optimized traffic flows for CAN. IEEE Transactions on Information Forensics and Security, 16, 601-616.

\bibitem{Nam2021Access}
Nam, M., Park, S., and Kim, D. S. (2021). Intrusion Detection Method Using Bi-Directional GPT for in-Vehicle Controller Area Networks. IEEE Access, 9, 124931-124944.

\bibitem{Ohira2021AsiaCCS}
Ohira, S., Desta, A. K., Arai, I., and Fujikawa, K. (2021, May). PLI-TDC: Super Fine Delay-Time Based Physical-Layer Identification with Time-to-Digital Converter for In-Vehicle Networks. In Proceedings of the 2021 ACM Asia Conference on Computer and Communications Security (pp. 176-186).

\bibitem{Xie2021TVT}
Xie, G., Yang, L. T., Liu, Y., Luo, H., Peng, X., and Li, R. (2021). Security enhancement for real-time independent in-vehicle CAN-FD messages in vehicular networks. IEEE Transactions on Vehicular Technology, 70(6).

\bibitem{Cover1991Elements}
Cover, T. M., and Thomas, J. A. (1991). Elements of information theory. John Wiley \& Sons.

\bibitem{Marchetti2016Evaluation}
Marchetti, M., Stabili, D., Guido, A., and Colajanni, M. (2016). Evaluation of anomaly detection for in-vehicle networks through information-theoretic algorithms. 2016 IEEE 2nd International Forum on Research and Technologies for Society and Industry Leveraging a Better Tomorrow (RTSI), 1–6. 

\bibitem{Wang2017Hardware}
Wang, E., Xu, W., Sastry, S., Liu, S., and Zeng, K. (2017). Hardware module-based message authentication in intra-vehicle networks. Proceedings of the 8th International Conference on Cyber-Physical Systems, 207–216. 

\bibitem{Groza2018TIFS}
Groza, B., and Murvay, P. S. (2018). Efficient intrusion detection with bloom filtering in controller area networks. IEEE Transactions on Information Forensics and Security, 14(4), 1037-1051.

\bibitem{Zhang2019TVT}
Zhang, L., Yang, F., and Lei, Y. (2019). Tree-based intermittent connection fault diagnosis for controller area network. IEEE Transactions on Vehicular Technology, 68(9), 9151-9161.

\bibitem{Liu2020ITS}
Liu, S., Zhang, H., Shao, L., and Yang, J. (2020). Built-in depth-semantic coupled encoding for scene parsing, vehicle detection and road segmentation. IEEE Transactions on Intelligent Transportation Systems, 22(9), 5520-5534.

\bibitem{Cordts2016CVPR}
Cordts, M., Omran, M., Ramos, S., Rehfeld, T., Enzweiler, M., Benenson, R., Franke, U., Roth, S. and Schiele, B. (2016). The cityscapes dataset for semantic urban scene understanding. In Proceedings of the 26th IEEE conference on computer vision and pattern recognition, 3213-3223.

\bibitem{Geiger2013JBR}
Geiger, A., Lenz, P., Stiller, C., and Urtasun, R. (2013). Vision meets robotics: The kitti dataset. The International Journal of Robotics Research, 32(11), 1231-1237.

\bibitem{Xie2021ITSOptimizing}
Xie, Y., Zeng, G., Kurachi, R., Xiao, F., and Takada, H. (2021). Optimizing Extensibility of CAN FD for Automotive Cyber-Physical Systems. IEEE Transactions on Intelligent Transportation Systems, 22(12), 7875-7886.

\bibitem{Song2020vehicle}
Song, H. M., Woo, J., and Kim, H. K. (2020). In-vehicle network intrusion detection using deep convolutional neural network. Vehicular Communications, 21, 100198.

\bibitem{VanWyk2019ITS}
Van Wyk, F., Wang, Y., Khojandi, A., and Masoud, N. (2019). Real-time sensor anomaly detection and identification in automated vehicles. IEEE Transactions on Intelligent Transportation Systems, 21(3), 1264-1276.

\bibitem{Yang2019Access}
Yang, Y., Wang, L., Li, Z., Shen, P., Guan, X., and Xia, W. (2019). Anomaly detection for controller area network in braking control system with dynamic ensemble selection. IEEE Access, 7, 95418-95429.

\bibitem{Javed2020ITS}
Javed, A. R., Usman, M., Rehman, S. U., Khan, M. U., and Haghighi, M. S. (2020). Anomaly detection in automated vehicles using multistage attention-based convolutional neural network. IEEE Transactions on Intelligent Transportation Systems, 22(7), 4291-4300.

\bibitem{Ashraf2020ITS}
Ashraf, J., Bakhshi, A. D., Moustafa, N., Khurshid, H., Javed, A., and Beheshti, A. (2020). Novel deep learning-enabled lstm autoencoder architecture for discovering anomalous events from intelligent transportation systems. IEEE Transactions on Intelligent Transportation Systems, 22(7), 4507-4518.

\bibitem{Moustafa2015MilCIS}
Moustafa, N., and Slay, J. (2015, November). UNSW-NB15: a comprehensive data set for network intrusion detection systems (UNSW-NB15 network data set). Proceedings of the 5th IEEE Military Communications and Information Systems Conference, 1-6.

\bibitem{Tariq2020CAS}
Tariq, S., Lee, S., Kim, H. K., and Woo, S. S. (2020). CAN-ADF: The controller area network attack detection framework. Computers \& Security, 94, 101857.

\bibitem{Islam2020ITS}
Islam, R., Refat, R. U. D., Yerram, S. M., and Malik, H. (2020). Graph-based intrusion detection system for controller area networks. IEEE Transactions on Intelligent Transportation Systems, 23(3), 1727-2736.

\bibitem{Moulahi2021Access}
Moulahi, T., Zidi, S., Alabdulatif, A., and Atiquzzaman, M. (2021). Comparative Performance Evaluation of Intrusion Detection Based on Machine Learning in In-Vehicle Controller Area Network Bus. IEEE Access, 9, 99595-99605.

\bibitem{Derhab2021ITS}
Derhab, A., Belaoued, M., Mohiuddin, I., Kurniawan, F., and Khan, M. K. (2021). Histogram-Based Intrusion Detection and Filtering Framework for Secure and Safe In-Vehicle Networks. IEEE Transactions on Intelligent Transportation Systems, 23(3), 2366-2379.

\bibitem{Liu2021TDSC}
Liu, J., and Park, J. (2021). ``Seeing is not Always Believing": Detecting Perception Error Attacks Against Autonomous Vehicles. IEEE Transactions on Dependable and Secure Computing, 18(5), 2209-2223.

\bibitem{Han2021TIFS}
Han, M. L., Kwak, B. I., and Kim, H. K. (2021). Event-Triggered Interval-Based Anomaly Detection and Attack Identification Methods for an In-Vehicle Network. IEEE Transactions on Information Forensics and Security, 16, 2941-2956.

\bibitem{Radu2016ESORICS}
Radu, A. I., and Garcia, F. D. (2016, September). LeiA: A lightweight authentication protocol for CAN. In European Symposium on Research in Computer Security (pp. 283-300). Springer, Cham.

\bibitem{Poudel2018TDSC}
Poudel, B., and Munir, A. (2018). Design and evaluation of a reconfigurable ecu architecture for secure and dependable automotive cps. IEEE Transactions on Dependable and Secure Computing.

\bibitem{Groza2019Access}
Groza, B., Popa, L., and Murvay, P. S. (2019). TRICKS—Time TRIggered Covert Key Sharing for Controller Area Networks. IEEE Access, 7, 104294-104307.

\bibitem{Joo2020NDSS}
Joo, K., Choi, W., and Lee, D. H. (2020). Hold the Door! Fingerprinting Your Car Key to Prevent Keyless Entry Car Theft. In NDSS.

\bibitem{Palaniswamy2020TIFS}
Palaniswamy, B., Camtepe, S., Foo, E., and Pieprzyk, J. (2020). An efficient authentication scheme for intra-vehicular controller area network. IEEE Transactions on Information Forensics and Security, 15, 3107-3122.

\bibitem{Jo2020TVT}
Jo, H. J., Kim, J. H., Choi, H. Y., Choi, W., Lee, D. H., and Lee, I. (2019). MAuth-CAN: Masquerade-Attack-Proof authentication for in-vehicle networks. IEEE transactions on vehicular technology, 69(2), 2204-2218.

\bibitem{Xiao2020ACSEC}
Xiao, Y., Shi, S., Zhang, N., Lou, W., and Hou, Y. T. (2020, December). Session Key Distribution Made Practical for CAN and CAN-FD Message Authentication. In Annual Computer Security Applications Conference, 681-693.

\bibitem{Xie2020ITSCANFD}
Xie, G., Yang, L. T., Wu, W., Zeng, K., Xiao, X., and Li, R. (2020). Security Enhancement for Real-Time Parallel In-Vehicle Applications by CAN FD Message Authentication. IEEE Transactions on Intelligent Transportation Systems.

\bibitem{Xiao2021TIFS}
Xiao, L., Lu, X., Xu, T., Zhuang, W., and Dai, H. (2021). Reinforcement Learning-Based Physical-Layer Authentication for Controller Area Networks. IEEE Transactions on Information Forensics and Security, 16, 2535-2547.

\bibitem{Plappert2021CCS}
Plappert, C., Jäger, L., and Fuchs, A. (2021, May). Secure Role and Rights Management for Automotive Access and Feature Activation. In Proceedings of the 2021 ACM Asia Conference on Computer and Communications Security (pp. 227-241).

\bibitem{Musuroi2021TVT}
Musuroi, A., Groza, B., Popa, L., and Murvay, P. S. (2021). Fast and Efficient Group Key Exchange in Controller Area Networks (CAN). IEEE Transactions on Vehicular Technology, 70(9), 9385-9399.

\bibitem{Ying2021TDSC}
Ying, X., Bernieri, G., Conti, M., Bushnell, L., and Poovendran, R. (2021). Covert Channel-Based Transmitter Authentication in Controller Area Networks. IEEE Transactions on Dependable and Secure Computing.

\bibitem{Limbasiya2022ICDCN}
Limbasiya, T., Ghosal, A., and Conti, M. (2022, January). AutoSec: Secure Automotive Data Transmission Scheme for In-Vehicle Networks. In 23rd International Conference on Distributed Computing and Networking (pp. 208-216).

\bibitem{Alvarez2014ITS}
Alvarez, J. M., López, A. M., Gevers, T., and Lumbreras, F. (2014). Combining priors, appearance, and context for road detection. IEEE Transactions on Intelligent Transportation Systems, 15(3), 1168-1178.

\bibitem{Jiang2019ICICS}
Jiang, J., Wang, C., Chattopadhyay, S., and Zhang, W. (2019, December). Road context-aware intrusion detection system for autonomous cars. In International Conference on Information and Communications Security (pp. 124-142). Springer, Cham.

\bibitem{Jo2015AVCarITIE}
Jo, K., Kim, J., Kim, D., Jang, C., and Sunwoo, M. (2015). Development of autonomous car—Part II: A case study on the implementation of an autonomous driving system based on distributed architecture. IEEE Transactions on Industrial Electronics, 62(8), 5119-5132.

\bibitem{Muter2010structured}
Muter, M., Groll, A., and Freiling, F. C. (2010). A structured approach to anomaly detection for in-vehicle networks. 2010 Sixth International Conference on Information Assurance and Security, 92–98. 

\bibitem{Rawat2021Decentralized}
Khodari, M., Rawat, A., Asplund, M., and Gurtov, A. (2019). Decentralized firmware attestation for in-vehicle networks. Proceedings of the 5th on Cyber-Physical System Security Workshop, 47–56.

\bibitem{Huang2018ATGARES}
Huang, T., Zhou, J., and Bytes, A. (2018, August). ATG: An attack traffic generation tool for security testing of in-vehicle CAN bus. In Proceedings of the 13th International Conference on Availability, Reliability and Security (pp. 1-6).

\bibitem{Choi2021TVTDBC}
Choi, W., Lee, S., Joo, K., Jo, H. J., and Lee, D. H. (2021). An Enhanced Method for Reverse Engineering CAN Data Payload. IEEE Transactions on Vehicular Technology, 70(4), 3371-3381.

\bibitem{Pham2019TSE}
Pham, V. T., Bohme, M., Santosa, A. E., Caciulescu, A. R., and Roychoudhury, A. (2019). Smart greybox fuzzing. IEEE Transactions on Software Engineering.

\bibitem{KLEEFuzzing}
KLEE Symbolic Execution Engine. [Online] \url{https://klee.github.io/}

\bibitem{RadamsaFuzzing}
How To Install and Use Radamsa to Fuzz Test Programs and Network Services on Ubuntu 18.04. [Online] \url{https://www.digitalocean.com/community/tutorials/how-to-install-and-use-radamsa-to-fuzz-test-programs-and-network-} \url{services-on-ubuntu-18-04}

\bibitem{Nishimura2016ICVES}
Nishimura, R., Kurachi, R., Ito, K., Miyasaka, T., Yamamoto, M., and Mishima, M. (2016, July). Implementation of the CAN-FD protocol in the fuzzing tool beSTORM. In 2016 IEEE International Conference on Vehicular Electronics and Safety (ICVES) (pp. 1-6). IEEE.

\bibitem{Sikder2021CSTSensorSurvey}
Sikder, A. K., Petracca, G., Aksu, H., Jaeger, T., and Uluagac, A. S. (2021). A Survey on Sensor-Based Threats and Attacks to Smart Devices and Applications. IEEE Communications Surveys \& Tutorials, 23(2), 1125-1159.

\bibitem{Ma2020JAS}
Ma, Y., Wang, Z., Yang, H., and Yang, L. (2020). Artificial intelligence applications in the development of autonomous vehicles: a survey. IEEE/CAA Journal of Automatica Sinica, 7(2), 315-329.

\bibitem{Zhu2022ITII6G}
Zhu, D., Bilal, M., and Xu, X. (2022). Edge Task Migration with 6G-Enabled Network in Box for Cybertwin based Internet of Vehicles. IEEE Transactions on Industrial Informatics, 18(7), 4893-4901.

\bibitem{Xu2022IoT6G}
Xu, X., Yao, L., Bilal, M., Wan, S., Dai, F., and Choo, K. K. R. (2022). Service migration across edge devices in 6G-enabled Internet of Vehicles networks. IEEE Internet of Things Journal, 9(3), 1930-1937.

\bibitem{Babun2021CNIoTSurvey}
Babun, L., Denney, K., Celik, Z. B., McDaniel, P., and Uluagac, A. S. (2021). A survey on IoT platforms: Communication, security, and privacy perspectives. Computer Networks, 192, 108040.

\bibitem{ElRewini2019VC}
El-Rewini, Z., Sadatsharan, K., Selvaraj, D. F., Plathottam, S. J., and Ranganathan, P. (2020). Cybersecurity challenges in vehicular communications. Vehicular Communications, 23, 100214.

\bibitem{Zhang2014IoTMalware}
Zhang, T., Antunes, H., and Aggarwal, S. (2014). Defending connected vehicles against malware: Challenges and a solution framework. IEEE Internet of Things journal, 1(1), 10-21.

\bibitem{Elkhail2021Access}
Elkhail, A. A., Refat, R. U. D., Habre, R., Hafeez, A., Bacha, A., and Malik, H. (2021). Vehicle Security: A Survey of Security Issues and Vulnerabilities, Malware Attacks and Defenses. IEEE Access, 9, 162401-162437.

\bibitem{Acharya2020IEEEAccess}
Acharya, S., Dvorkin, Y., Pandzic, H., and Karri, R. (2020). Cybersecurity of smart electric vehicle charging: A power grid perspective. IEEE Access, 8, 214434-214453.

\bibitem{Wang2020DCN}
Wang, M., Zhu, T., Zhang, T., Zhang, J., Yu, S., and Zhou, W. (2020). Security and privacy in 6G networks: New areas and new challenges. Digital Communications and Networks, 6(3), 281-291.

\bibitem{Mughal2020VC}
Mughal, U. A., Xiao, J., Ahmad, I., and Chang, K. (2020). Cooperative resource management for C-V2I communications in a dense urban environment. Vehicular Communications, 26, 100282.

\bibitem{Nguyen2015CVPR}
Nguyen, A., Yosinski, J., and Clune, J. (2015). Deep neural networks are easily fooled: High confidence predictions for unrecognizable images. In Proceedings of the IEEE conference on computer vision and pattern recognition (pp. 427-436).

\bibitem{Qayyum2020IEEECOMST}
Qayyum, A., Usama, M., Qadir, J., and Al Fuqaha, A. (2020). Securing connected \& autonomous vehicles: Challenges posed by adversarial machine learning and the way forward. IEEE Communications Surveys \& Tutorials, 22(2), 998-1026.

\bibitem{Chen2019CybersecurityAI}
Chen, T., Liu, J., Xiang, Y., Niu, W., Tong, E., and Han, Z. (2019). Adversarial attack and defense in reinforcement learning-from AI security view. Cybersecurity, 2(1), 1-22.

\bibitem{Kong2020CVPR}
Kong, Z., Guo, J., Li, A., and Liu, C. (2020). Physgan: Generating physical-world-resilient adversarial examples for autonomous driving. In Proceedings of the IEEE/CVF Conference on Computer Vision and Pattern Recognition (pp. 14254-14263).

\bibitem{Nassi2020CCS}
Nassi, B., Mirsky, Y., Nassi, D., Ben-Netanel, R., Drokin, O., and Elovici, Y. (2020, October). Phantom of the ADAS: Securing advanced driver-assistance systems from split-second phantom attacks. In Proceedings of the 2020 ACM SIGSAC Conference on Computer and Communications Security (pp. 293-308).

\bibitem{Wang2021CCS}
Wang, W., Yao, Y., Liu, X., Li, X., Hao, P., and Zhu, T. (2021, November). I Can See the Light: Attacks on Autonomous Vehicles Using Invisible Lights. In Proceedings of the 2021 ACM SIGSAC Conference on Computer and Communications Security (pp. 1930-1944).

\bibitem{Deng2020Percom}
Deng, Y., Zheng, X., Zhang, T., Chen, C., Lou, G., and Kim, M. (2020, March). An analysis of adversarial attacks and defenses on autonomous driving models. In 2020 IEEE international conference on pervasive computing and communications (PerCom) (pp. 1-10). IEEE.

\bibitem{Hussain2018CSAT}
Hussain, R., and Zeadally, S. (2018). Autonomous cars: Research results, issues, and future challenges. IEEE Communications Surveys \& Tutorials, 21(2), 1275-1313.

\bibitem{Pereira2019FGCS}
Pereira, J., Ricardo, L., Luis, M., Senna, C., and Sargento, S. (2019). Assessing the reliability of fog computing for smart mobility applications in VANETs. Future Generation Computer Systems, 94, 317-332.

\bibitem{Sun2019JNCA}
Sun, G., Sun, S., Sun, J., Yu, H., Du, X., and Guizani, M. (2019). Security and privacy preservation in fog-based crowd sensing on the internet of vehicles. Journal of Network and Computer Applications, 134, 89-99.

\bibitem{Ni2017ICM}
Ni, J., Zhang, A., Lin, X., and Shen, X. S. (2017). Security, privacy, and fairness in fog-based vehicular crowdsensing. IEEE Communications Magazine, 55(6), 146-152.

\bibitem{Kang2017ITS}
Kang, J., Yu, R., Huang, X., and Zhang, Y. (2017). Privacy-preserved pseudonym scheme for fog computing supported internet of vehicles. IEEE Transactions on Intelligent Transportation Systems, 19(8), 2627-2637.

\bibitem{Nkenyereye2021}
Nkenyereye, L., Islam, S. R., Bilal, M., Abdullah-Al-Wadud, M., Alamri, A., and Nayyar, A. (2021). Secure crowd-sensing protocol for fog-based vehicular cloud. Future Generation Computer Systems, 120, 61-75.

\bibitem{Soleymani2021VC}
Soleymani, S. A., Goudarzi, S., Anisi, M. H., Zareei, M., Abdullah, A. H., and Kama, N. (2021). A security and privacy scheme based on node and message authentication and trust in fog-enabled VANET. Vehicular Communications, 29, 100335.

\bibitem{Liu2021MNAA}
Liu, L., Chen, C., Pei, Q., Maharjan, S., and Zhang, Y. (2021). Vehicular edge computing and networking: A survey. Mobile networks and applications, 26(3), 1145-1168.

\bibitem{Singh2021JPDC}
Singh, J., Singh, P., and Gill, S. S. (2021). Fog computing: A taxonomy, systematic review, current trends and research challenges. Journal of Parallel and Distributed Computing, 157, 56-85.

\bibitem{Wu2021IoT}
Wu, Y., Zhang, K., and Zhang, Y. (2021). Digital twin networks: A survey. IEEE Internet of Things Journal, 8(18), 13789-13804.

\bibitem{Jones2020CIRP}
Jones, D., Snider, C., Nassehi, A., Yon, J., and Hicks, B. (2020). Characterising the Digital Twin: A systematic literature review. CIRP Journal of Manufacturing Science and Technology, 29, 36-52.

\bibitem{Li2021TVT}
Li, G., Lai, C., Lu, R., and Zheng, D. (2021). Seccdv: A security reference architecture for cybertwin-driven 6g v2x. IEEE Transactions on Vehicular Technology.

\bibitem{Guo2022DCN}
Guo, J., Bilal, M., Qiu, Y., Qian, C., Xu, X., and Choo, K. K. R. (2022). Survey on digital twins for Internet of Vehicles: Fundamentals, challenges, and opportunities. Digital Communications and Networks, 1-17.

\bibitem{Alcaraz2022CSAT}
Alcaraz, C., and Lopez, J. (2022). Digital Twin: A Comprehensive Survey of Security Threats. IEEE Communications Surveys \& Tutorials, pp. 1-29.

\end{thebibliography}
\end{document}